\documentclass[11pt]{elsart}
\usepackage{graphicx,color,epsfig,amsmath,amssymb}
\usepackage{subcaption}
\usepackage{cite,hyperref}
\usepackage{listings}
\usepackage[super]{nth}
\graphicspath{{plots/}}
\newcounter{bla}
\newenvironment{refnummer}{%
\list{[\arabic{bla}]}%
{\usecounter{bla}%
 \setlength{\itemindent}{0pt}%
 \setlength{\topsep}{0pt}%
 \setlength{\itemsep}{0pt}%
 \setlength{\labelsep}{2pt}%
 \setlength{\listparindent}{0pt}%
 \settowidth{\labelwidth}{[9]}%
 \setlength{\leftmargin}{\labelwidth}%
 \addtolength{\leftmargin}{\labelsep}%
 \setlength{\rightmargin}{0pt}}}
 {\endlist}
\def\be{\begin{align}}
\def\ee{\end{align}}
\def\bea{\begin{align}}
\def\eea{\end{align}}
\def\nn{\nonumber}

\newcommand{\pysecdec}{py{\textsc{SecDec}}}

\newcommand{\python}{{\texttt{python}}}

\newcommand{\eps}{\epsilon}

\newcommand{\cppeleven}{{\texttt{c++11}}}
\newcommand{\cpp}{{\texttt{c++}}}
\newcommand{\cuda}{{\textsc{CUDA}}}
\newcommand{\cuba}{{\textsc{Cuba}}}
\newcommand{\vegas}{{\textsc{Vegas}}}
\newcommand{\suave}{{\textsc{Suave}}}
\newcommand{\cuhre}{{\textsc{Cuhre}}}
\newcommand{\divonne}{{\textsc{Divonne}}}
\newcommand{\qmc}{{QMC}}

\begin{document}

\begin{frontmatter}
\hfill{MPP-2018-279, CERN-TH-2018-246, ZU-TH 43/18, IPPP/18/100}\\ 

%
\title{A GPU compatible quasi-Monte Carlo integrator interfaced to pySecDec}

\author[a]{S.~Borowka},
\author[b]{G.~Heinrich},
\author[b]{S.~Jahn},
\author[a,b]{S.~P.~Jones},
\author[b,c]{M.~Kerner},
\author[d]{J.~Schlenk}

\address[a]{Theoretical Physics Department, CERN, Geneva, Switzerland}
\address[b]{Max Planck Institute for Physics, F\"ohringer Ring 6, 80805 M\"unchen, Germany}
\address[c]{Physik-Institut, Universit{\"a}t Z{\"u}rich, Winterthurerstrasse 190, 8057 Z{\"u}rich, Switzerland}
\address[d]{Institute for Particle Physics Phenomenology, University of Durham, Durham DH1 3LE, UK}

\begin{abstract}
The purely numerical evaluation of multi-loop integrals and amplitudes
can be a viable alternative to analytic approaches, in particular in
the presence of several mass scales, provided sufficient accuracy 
can be achieved in an acceptable amount of time.
For many multi-loop integrals, the fraction of time required to perform 
the numerical integration is significant and it is therefore beneficial to have efficient and well-implemented numerical integration methods.
With this goal in mind, we present a new stand-alone integrator based on the use of (quasi-Monte Carlo) rank-1 shifted lattice rules.
For integrals with high variance we also implement a variance reduction algorithm based on fitting a smooth function to the inverse cumulative distribution function of the integrand dimension-by-dimension.

Additionally, the new integrator is interfaced to \pysecdec{} to allow the straightforward evaluation of multi-loop integrals and dimensionally regulated parameter integrals.
In order to make use of recent advances in parallel computing hardware, our integrator can be used both on CPUs and \cuda{} compatible GPUs where available.
\end{abstract}


\begin{keyword}
Perturbation theory, Feynman diagrams, multi-loop, numerical integration
\end{keyword}

\end{frontmatter}

\newpage

{\bf PROGRAM SUMMARY}
   
\begin{small}
\noindent
{\em Manuscript Title: } A GPU compatible quasi-Monte Carlo integrator interfaced to pySecDec   \\
{\em Authors: } S.~Borowka, G.~Heinrich, S.~Jahn, S.~P.~Jones, M.~Kerner, J.~Schlenk   \\
{\em Program Title: } pySecDec, qmc                                        \\
{\em Journal Reference:}    https://doi.org/10.1016/j.cpc.2019.02.015                                  \\
 {\em Licensing provisions: GNU Public License v3}                                   \\
{\em Programming language:} python, FORM, C++, CUDA     \\
{\em Computer:} from a single PC/Laptop to a cluster, depending on the
problem; if the optional GPU support is used, CUDA compatible hardware
with Compute Capability 3.0 or greater is required.\\
{\em Operating system: } Unix, Linux                                      \\
{\em RAM:} depending on the complexity of the problem                                              \\
{\em Keywords:}  Perturbation theory, Feynman diagrams, multi-loop, numerical integration
 \\
{\em Classification:}                                         
  4.4 Feynman diagrams, 
  5 Computer Algebra, 
  11.1 General, High Energy Physics and Computing.\\
 {\em External routines/libraries:} 
catch [1],
gsl [2],
numpy [3],
sympy [4],
Nauty [5],
Cuba [6],  
FORM [7], 
Normaliz [8]. 
The program can also be used in a mode which does not require Normaliz.   \\
{\em Journal reference of previous version:} Comput. Phys. Commun. 222 (2018) 313-326. \\
{\em Does the new version supersede the previous version?:} yes  \\
{\em Nature of the problem:}\\
  Extraction of ultraviolet and infrared singularities from parametric integrals 
  appearing in higher order perturbative calculations in quantum field
  theory. 
  Numerical integration in the presence of integrable singularities 
  (e.g. kinematic thresholds). \\
{\em Solution method:}\\
 Algebraic extraction of singularities within dimensional regularization using iterated sector decomposition. 
 This leads to a Laurent series in the dimensional regularization
 parameter $\epsilon$
 (and optionally other regulators), 
 where the coefficients are finite integrals over the unit-hypercube. 
 Those integrals are evaluated numerically by Monte Carlo integration.
 The integrable singularities are handled by 
 choosing a suitable integration contour in the complex plane, in an
 automated way.
 The parameter integrals forming the coefficients of the Laurent
 series in the regulator(s) are provided in the form of libraries
 which can be linked to the calculation of (multi-) loop amplitudes.
   \\
{\em Restrictions:} Depending on the complexity of the problem, limited by 
memory and CPU/GPU time.\\
{\em Running time:}
Between a few seconds and several days, depending on the complexity of the problem.\\
{\em References:}
\begin{refnummer}
\item https://github.com/philsquared/Catch/.
\item http://www.gnu.org/software/gsl/.
\item http://www.numpy.org/.
\item http://www.sympy.org/.
\item http://pallini.di.uniroma1.it/.
\item T.~Hahn, 
``CUBA: A Library for multidimensional numerical integration,''
  Comput.\ Phys.\ Commun.\  {\bf 168} (2005) 78 
  [hep-ph/0404043], 
http://www.feynarts.de/cuba/.
\item J.~Kuipers, T.~Ueda and J.~A.~M.~Vermaseren,
  ``Code Optimization in FORM,''
  Comput.\ Phys.\ Commun.\  {\bf 189} (2015) 1
  [arXiv:1310.7007], 
http://www.nikhef.nl/~form/.
\item W.~Bruns, B.~Ichim, B. and T.~R{\"o}mer, C.~S{\"o}ger, 
``Normaliz. Algorithms for rational cones and affine monoids.''
 http://www.math.uos.de/normaliz/.
\end{refnummer}
\end{small}


\section{Introduction}
\label{sec:intro}

High energy particle physics is in an era where the current
underlying theory, the Standard Model (SM), is very well tested
experimentally, as well as
consistent and therefore predictive from a theoretical point of view. 
This means that we can control the SM predictions very well, and so 
should be able -- at least in principle -- to identify physics beyond the SM
even if it is showing up only in small deviations. 

In practice, there are several obstacles when trying to increase the
precision of theoretical predictions. 
Focusing on problems accessible to perturbation theory, a major
obstacle is the fast increase in complexity of the calculation as the
number of loops and the number of kinematic scales increases.
Despite the remarkable progress that has been achieved in the analytic
calculation of multi-loop amplitudes and integrals in the last few
years, analytical approaches are only at the beginning of a journey
into largely unexplored mathematical territory if the function class
of the results goes beyond multiple polylogarithms (MPLs), typically
involving elliptic or hyper-elliptic functions,
see
e.g.~\cite{Laporta:2004rb,Adams:2017tga,Abreu:2017enx,Primo:2017ipr,Bourjaily:2017bsb,Broedel:2017kkb,Adams:2018yfj,Broedel:2018iwv,Lee:2018ojn,Bourjaily:2018ycu,Blumlein:2018cms,Broedel:2018qkq,Bourjaily:2018yfy}.
 
On the other hand, (semi-)numerical approaches do not necessarily become less efficient
if the result leaves the class of MPLs.
This is one of the reasons why it is important to develop numerical
methods which are fast and accurate enough to provide results where
analytic approaches are at their limits.
Sector decomposition~\cite{Hepp:1966eg,Roth:1996pd,Binoth:2000ps,Heinrich:2008si}
is an example of such a method;  other recent semi-numerical methods
are described e.g. in Refs.~\cite{Becker:2012bi,Sborlini:2016hat,Freitas:2016sty,deDoncker:2017gnb,Gluza:2016fwh,Usovitsch:2018shx,Baglio:2018lrj,Bendavid:2018nar,Blondel:2018mad}.
Sector decomposition is a procedure which can be applied to dimensionally regulated
integrals in order to factorise singularities in the regulator. The resulting finite parameter integrals, 
which form the coefficients at each order in the regulator, can then be numerically integrated.  
There are several public
implementations of sector decomposition~\cite{Bogner:2007cr,Gluza:2010rn,Ueda:2009xx,Smirnov:2008py,Smirnov:2009pb,Smirnov:2013eza,Smirnov:2015mct,Borowka:2015mxa,Borowka:2017idc}.
Recently, an analytical method,
based on sector decomposition followed by a series expansion in the Feynman parameters and analytic integration, has
been worked out in Ref.~\cite{Borowka:2018dsa}.

Currently, in the publicly available sector decomposition tools, numerical integration mostly relies on either deterministic
integration rules for integrals of low dimensionality or Monte Carlo
integration, as implemented in the {\sc Cuba}
library~\cite{Hahn:2004fe,Hahn:2014fua}.
However, the integration error for Monte Carlo integration scales only like $1/\sqrt{n}$, where $n$ is
the number of samples, which limits the accuracy that can be obtained in a given integration time. 
To improve on this, a different integration
method has to be chosen. One such method is the {\it quasi-Monte
Carlo} (\qmc{}) method~\cite{QMCActaNumerica}, where the integration
error scales like  $1/n$ or better, rather than $1/\sqrt{n}$.
In order for this scaling to be achieved, the integrand functions need
to fulfil certain requirements. The \qmc{} method discussed herein was first applied to functions
produced by the sector decomposition algorithm in
Ref.~\cite{Li:2015foa},
where it was shown practically that the conditions for $1/n$ or better scaling are usually met and 
the good performance of Graphics Processing Units (GPUs) when evaluating such functions was also demonstrated.
An application of quasi-Monte Carlo methods to two- and three-loop integrals also has been presented in Ref.~\cite{deDoncker:2018nqe}.
The \qmc{} method, implemented to run on GPUs, has already been applied successfully 
to phenomenological applications involving multi-scale two-loop
integrals including Higgs-boson pair production~\cite{Borowka:2016ehy,Borowka:2016ypz} and H+jet production~\cite{Jones:2018hbb} at NLO.

In this work, we present a new stand-alone \qmc{} integrator capable of utilising multiple cores of Central Processing Units (CPUs)
and multiple Graphics Processing Units (GPUs). We also present a new version of the program \pysecdec{}
which makes available our \qmc{} implementation as an additional integrator.
Furthermore, we present and implement a method for combining the \qmc{} integration with importance sampling.
We emphasize that our \qmc{} implementation can also be straightforwardly used outside of the
\pysecdec{} context.

The outline of the paper is as follows. In Section \ref{sec:mathintro} 
we give an overview on the \qmc{} method as implemented in our
program and describe our variance reduction procedure.
Section \ref{sec:standalone} is dedicated to the stand-alone usage of
the \qmc{} integrator library, we also describe the design of the library and some basics regarding the use of GPUs.
In Section \ref{sec:pysecdec} we explain the usage of the \qmc{} integrator within \pysecdec{} and describe various examples.
Section \ref{sec:profiling} is dedicated to profiling the \qmc{} method and our implementation.
After we conclude in Section \ref{sec:conclusion}, we provide detailed API documentation in Appendix \ref{sec:appendix:api}.

\section{Description of the \qmc{} method}
\label{sec:mathintro}
\subsection{Quasi-Monte Carlo integration}


Our aim is to numerically compute the multiple integral of a function $f : \mathbb{R}^d \rightarrow \mathbb{R}$ or $f : \mathbb{R}^d \rightarrow \mathbb{C}$ over a $d$-dimensional unit hypercube $[0,1]^d$,
\begin{equation}
I[ f ] \equiv \int_{[0,1]^d} \mathrm{d} \mathbf{x} \ f(\mathbf{x}) \equiv \int_0^1 \mathrm{d} x_1 \cdots \mathrm{d} x_d \ f(x_1,\ldots,x_d)\;.
\end{equation}

In this section we will briefly introduce the concept of quasi-Monte Carlo integration
and state the most relevant results and formulae. The study of \qmc{} integration
has produced a vast amount of literature, for a more thorough review we refer the reader to the existing mathematical literature,
for example Refs.~\cite{QMCActaNumerica,KuoNuyensPractical} and references therein.

Unlike Monte Carlo integrators, quasi-Monte Carlo (\qmc{}) 
integrators are based on a predominantly deterministic numerical integration. 
An unbiased estimate $\bar{Q}_{n,m}[f]$ of the integral $I[f]$ can be obtained from the following (\qmc{}) cubature rule, 
known as a rank-1 shifted lattice (R1SL) rule~\cite{QMCActaNumerica}:
\begin{align}
&I[f] \approx \bar{Q}_{n,m}[f] \equiv  \frac{1}{m} \sum_{k=0}^{m-1} Q_{n}^{(k)}[f], &
&Q_{n}^{(k)}[f] \equiv \frac{1}{n} \sum_{i=0}^{n-1} f \left( \left\{ \frac{i \mathbf{z}}{n} + \boldsymbol{\Delta}_k \right\} \right)\;.& \label{eq:lattice}
\end{align}
The rank of the rule denotes the minimal number of generating vectors required to generate the lattice rule. In this work we will consider only rank-1 lattices i.e. those generated by a single generating vector.
The estimate of the integral depends on the number of lattice points $n$ and the number of random shifts $m$. The shift vectors $\boldsymbol{\Delta}_k \in [0,1)^d$ are $d$-dimensional vectors with components consisting of independent, uniformly distributed random real numbers in the interval $[0,1)$. The generating vector $\mathbf{z} \in \mathbb{Z}^d$ is a fixed $d$-dimensional vector of integers coprime to $n$. The curly brackets indicate that the fractional part of each component is taken, such that all arguments of $f$ remain in the interval $[0,1)$. 

A reliable estimate of the integral can be obtained even without random shifts provided that the lattice is sufficiently large, however, the random shifts allow the remaining error to be estimated. More precisely,  an unbiased estimate of the mean-square error can be obtained from the random shifts of the lattice according to
\begin{equation}
\sigma_{n,m}^2[f] \equiv \mathrm{Var}[\bar{Q}_{n,m}[f]] \approx \frac{1}{m(m-1)} \sum_{k=0}^{m-1} ( Q_{n}^{(k)}[f] - \bar{Q}_{n,m}[f] )^2\;. \label{eq:lattice-error}
\end{equation}
In typical applications only 10-20 random shifts are required to obtain a reliable estimate of the error.

\begin{figure}
\begin{center}
\includegraphics[width=0.85\textwidth]{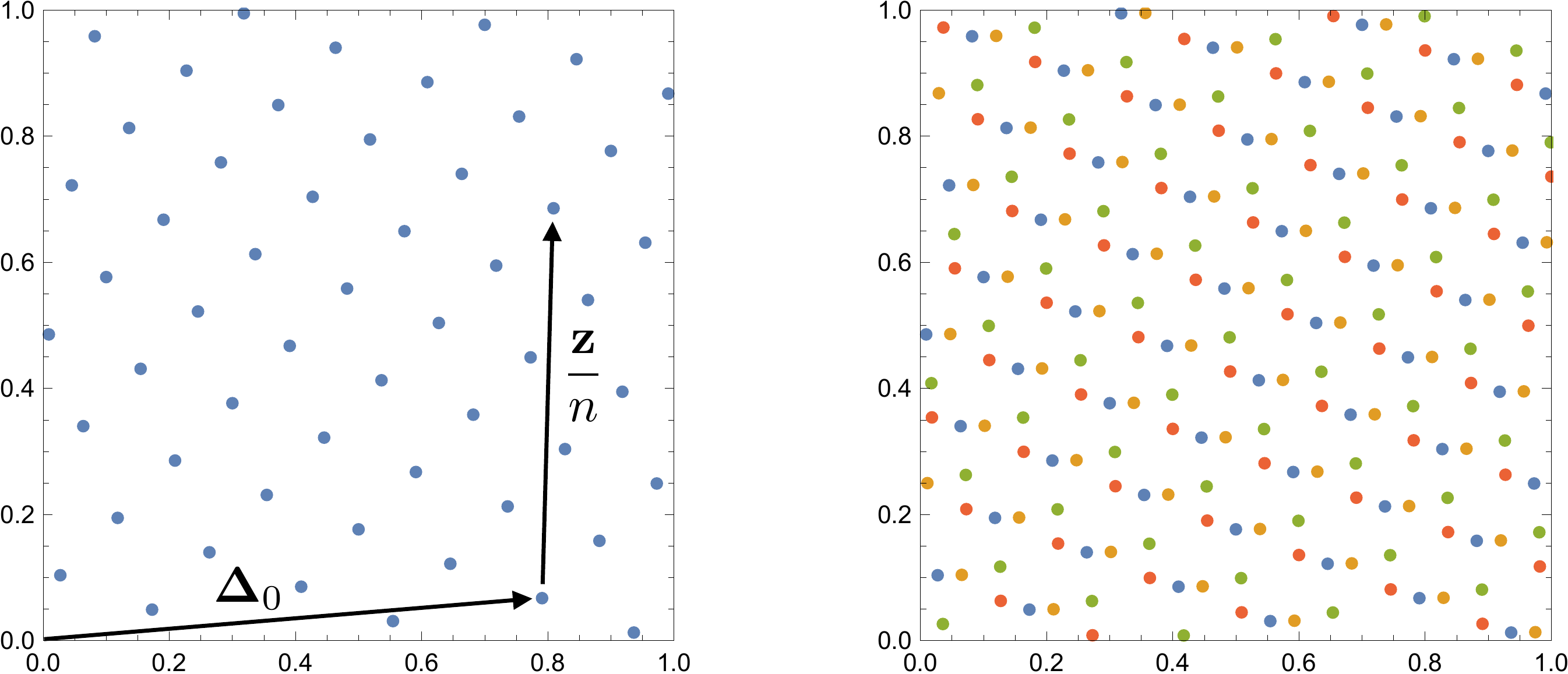}
\end{center}
\caption{(Left panel) A $d=2$ dimensional R1SL with $n=55$ points, generating vector $\mathbf{z}=(1,34)$ and random shift $\boldsymbol{\Delta}_0$. (Right panel) A R1SL produced with three additional random shifts, which can used to estimate the mean-square error as described in the text.}
\label{fig:lattice-example}
\end{figure}

In Figure~\ref{fig:lattice-example} an example shifted lattice is shown. In the left panel a single lattice is displayed. The zeroth point is shifted from the origin by the random shift vector $\boldsymbol{\Delta}_0$. Further points are generated by adding $\mathbf{z}/n$ and wrapping back into the unit square as necessary. The lattice displayed contains a total of $n=55$ points. In the right panel, three additional shifted lattices are displayed. They are generated by shifting the original lattice and can be used to produce an estimate of the integration error using Eq.~\ref{eq:lattice-error}.

The classical theoretical error bounds on \qmc{} rules take the form of a product
\begin{equation}
|I[f]-Q_n[f]| \le D(\mathbf{t}_0, \ldots, \mathbf{t}_{n-1}) V[f],
\end{equation}
where $\mathbf{t}_i$ are the cubature points generated by the generating vector(s), $D$ is the discrepancy of the point set and $V$ is the variation of $f$. The discrepancy depends only on the points and the variation depends only on the integrand. If $f$ can be differentiated once with respect to each variable then it can be proven that for a particular choice of cubature points (or, equivalently, a particular generating vector) \qmc{} methods converge as $\mathcal{O}((\log n)^d/n)$. This error bound grows exponentially with dimension, seemingly implying that \qmc{} integration is not useful in a large number of dimensions.

However, by working with weighted function spaces, it can be shown that the error bound can be independent of the dimension provided that the variables of the integrand $f$ have some varying degree of importance. In the modern literature, error bounds have been studied in terms of the product
\begin{equation}
|I[f]-Q_n[f]| \le e_{\boldsymbol{\gamma}}(\mathbf{t}_0, \ldots, \mathbf{t}_{n-1}) ||f||_{\boldsymbol{\gamma}}, \label{eq:qmcerrorproduct}
\end{equation} 
where $e_{\boldsymbol{\gamma}}$ is the worst case error in a weighted function space with weights $\boldsymbol{\gamma}$ and $||f||_{\boldsymbol{\gamma}}$ is the norm of $f$ in the weighted space.

Following Ref.~\cite{QMCActaNumerica} we will discuss two function spaces (Sobolev spaces and Korobov spaces) that allow important properties of the \qmc{} to be proven. In both cases we will state theorems from the literature that bound the worst case error for a rank-1 shifted lattice rule in the corresponding function space. By definition, the worst case error for a shifted lattice rule in the weighted function space is the largest possible error for any function with norm less than or equal to 1,
\begin{equation}
e_{\boldsymbol{\gamma}}(\mathbf{z} ,\boldsymbol{\Delta}) \equiv \sup_{||f||_{\boldsymbol{\gamma}} \le 1} | I[f] - Q_n[f]|.
\end{equation}
Here we use the notation $e_{\boldsymbol{\gamma}}(\mathbf{z} ,\boldsymbol{\Delta})$ in place of $e_{\boldsymbol{\gamma}}(\mathbf{t}_0, \ldots, \mathbf{t}_{n-1})$ to refer to the worst case error of the point set generated by $\mathbf{z}$ (the generating vector) and $\boldsymbol{\Delta}$ (the random shift vector).
The shift averaged worst case error, $e^\mathrm{sh}_{\boldsymbol{\gamma}}(\mathbf{z})$, is given by averaging the worst case error over uniformly distributed shifts in $[0,1]^d$. The choice of weights, $\boldsymbol{\gamma}$, affects both the norm of the function and the worst case error. Choosing large weights leads to a smaller norm but larger worst case error and vice versa.

First we consider a Sobolev space spanned by functions $f$ with square integrable (weak) derivatives $\frac{\partial^{|\mathbf{a}|}f(\mathbf{x})}{\partial \mathbf{x}_{\mathbf{a}}}$ and $\mathbf{a} \in \{0,1\}^d$. The norm of $f$ in the weighted Sobolev space can be written as
\begin{equation}
||f||^2_{\mathrm{Sobolev},\boldsymbol{\gamma}} = \sum_{\mathfrak{u} \subseteq \{1,\ldots,d\} } \frac{1}{\gamma_\mathfrak{u}} \int_{[0,1]^{|\mathfrak{u}|}} \left( \int_{[0,1]^{d-|\mathfrak{u}|}} \frac{\partial^{|\mathfrak{u}|} f(\mathbf{x})}{\partial \mathbf{x}_\mathfrak{u}} \mathrm{d}\mathbf{x}_{-\mathfrak{u}} \right)^2 \mathrm{d} \mathbf{x}_\mathfrak{u}.
\end{equation}
Note that the norm depends only on the mixed \emph{first} derivative because we never differentiate more than once with respect to a particular variable. It can be shown~\cite{QMCActaNumerica,SLOAN19981} that for functions belonging to such a space a R1SL rule exists for which the shift averaged worst case error is given by
\begin{equation}
  [e_\mathrm{Sobolev}^\mathrm{sh}(\mathbf{z})]^2 \le \left( \frac{1}{\psi(n)} \sum_{\emptyset \neq \mathfrak{u} \subseteq \{1,\ldots,d\}} \gamma_\mathfrak{u}^\lambda \left( \frac{2 \zeta(2\lambda)}{(2 \pi^2)^\lambda} \right)^{|\mathfrak{u}|} \right)^\frac{1}{\lambda}\label{eq:errorSobolev}
\end{equation}
for all $\lambda \in (1/2,1]$. Here $\zeta$ is the Riemann zeta function and $\psi$ is the Euler totient function. This formula indicates that, for suitably chosen weights, R1SL rules can have a convergence rate close to $\mathcal{O}(n^{-1})$ independently of $d$ for functions belonging to a Sobolev space.

The Korobov function space is a space of periodic functions which are $\alpha$ times differentiable in each variable. The parameter $\alpha$ is known as the smoothness parameter and characterises the rate of decay of the Fourier coefficients of the integrand. The norm of $f$ in the weighted Korobov space is given by
\begin{equation}
||f||^2_{\mathrm{Korobov},\boldsymbol{\gamma}} = \sum_{\mathbf{h} \in \mathbb{Z}^d} \frac{\prod_{j \in \mathfrak{u}(\mathbf{h})} |h_j|^{2\alpha}}{\gamma_{\mathfrak{u}(\mathbf{h})}}    |\hat{f}(\mathbf{h})|^2,
\end{equation}
where $\mathfrak{u}(\mathbf{h}) := \{ j \in \{1,\ldots,d \} : h_j \neq 0\}$ and $\hat{f}(\mathbf{h})$ are the Fourier coefficients of the integrand, given by
\begin{equation}
  \hat{f}(\mathbf{h}) = \int_{[0,1]^d} f(\mathbf{x}) e^{-2 \pi i \mathbf{h}\cdot \mathbf{x}} \mathrm{d} \mathbf{x}.\label{eq:errorKorobov}
\end{equation}
For functions belonging to a Korobov space with smoothness $\alpha$ the shift averaged worst case error is given by
\begin{equation}
[e_\mathrm{Korobov}^\mathrm{sh}(\mathbf{z})]^2 \le \left( \frac{1}{\psi(n)} \sum_{\emptyset \neq \mathfrak{u} \subseteq \{1,\ldots,d\}} \gamma_\mathfrak{u}^\lambda \left( 2 \zeta(2 \alpha \lambda) \right)^{|\mathfrak{u}|} \right)^\frac{1}{\lambda},
\end{equation}
for all $\lambda \in (1/(2\alpha),1]$. The best convergence rate is obtained when $\lambda \rightarrow 1/(2\alpha)$, which yields a convergence close to $\mathcal{O}(n^{-\alpha})$ independently of $d$ (for suitably chosen weights). Functions which are smooth but not periodic can be periodized by an integral transform as described in Section~\ref{sec:transformations}. This can improve the rate of convergence of quasi-Monte Carlo integration but may also increase the variance (or norm) of the function, especially in high dimensions.

\subsection{Generating vectors}
The convergence of the rank-1 lattice rule given in Eq.~\eqref{eq:lattice} depends on the choice of the generating vector $\mathbf z$ and in particular the worst case errors given in Eqs.~\eqref{eq:errorSobolev} and~\eqref{eq:errorKorobov} can only be achieved with specific choices of $\mathbf z$. An efficient algorithm to construct good generating vectors $\mathbf z$ is the component-by-component construction~\cite{nuyens2006fast}, where a generating vector in $d$ dimensions is obtained from a $d-1$ dimensional one by selecting the additional component such that the worst-case error is minimal. This allows to construct the generating vectors with a cost of $\mathcal O(d\,n\log n)$.

We provide generating vectors for different fixed lattice sizes $n$, which have been obtained for a Korobov space with product weights $\gamma_\mathfrak{u} = \prod_{i\in\mathfrak{u}} \gamma_i$, where we set all weights equal, $\gamma_i = 1/d$. 
More details on the generating vectors provided with the \qmc{} library are given in Appendix~\ref{sec:appendix:generatingVectors}.

\subsection{Transformations}
\label{sec:transformations}

Lattice rules perform particularly well for continuous and smooth functions which are periodic with respect to each variable. 
Sector decomposed functions are typically continuous and smooth but not periodic. 
However, they can be periodized by a suitable change of variables
$\mathbf{x}=\phi(\mathbf{u})$,
\begin{equation}
I[ f ] \equiv \int_{[0,1]^d} \mathrm{d} \mathbf{x} \ f(\mathbf{x}) = \int_{[0,1]^d} \mathrm{d} \mathbf{u} \ \omega_d(\mathbf{u}) f(\phi({\mathbf{u})})
\end{equation}
where 
\begin{align} 
&\phi(\mathbf{u})= (\phi(u_1),\ldots,\phi(u_d)), & &  \omega_d(\mathbf{u}) =  \prod_{j=1}^d \omega(u_j) & & \mathrm{and} & & \omega(u) = \phi^\prime(u).&
\end{align}
In practice, the periodizing transform may be specified in terms of the weight function, $\omega$, in which case the change of variables is given by
\begin{equation}
\phi(u) \equiv \int_0^u \mathrm{d} t \ \omega(t).
\end{equation}

We have implemented the following periodizing transformations:
\begin{itemize}
\item Korobov transforms~\cite{Korobov1963,LAURIE1996337,Kuo2007},
\item Sidi transforms~\cite{Sidi1993},
\item Baker's transform~\cite{baker_trafo}.
\end{itemize}
The Korobov transform is defined by the polynomial weight function
\begin{align}
&\omega_{r_0,r_1}(u)= \frac{u^{r_0}(1-u)^{r_1}}{\int_0^1 \mathrm{d} t \  t^{r_0} (1-t)^{r_1}} = (r_0+r_1+1) \binom{r_0+r_1}{r_0} u^{r_0} (1-u)^{r_1}, \label{eq:korobov}
\end{align}
The weight parameters $r_0,r_1$ are usually chosen to be equal. 
The behaviour of the integrand near the endpoints should be taken into account when choosing the weight parameters $r$, 
as the variance of the integral can depend critically on their choice.
Asymmetric Korobov transforms with $r_0\not=r_1$ can be beneficial in cases where the integrand approaches a singularity near one of the endpoints, while the other endpoint does not exhibit any singular behaviour.

Sidi transforms~\cite{Sidi1993,LAURIE1996337} are trigonometric integral transforms with a weight proportional to $(\sin \pi u)^r$:
\begin{align}
\omega_{r}(u) = \frac{(\sin{\pi u})^r}{\int_0^1\mathrm{d} t \ (\sin{\pi t})^r} = \frac{\pi}{2^r} \frac{\Gamma(r+1)}{\Gamma((r+1)/2)^2} (\sin \pi u)^r .
\label{eq:sidi}
\end{align}
The Sidi transforms may be used to periodize an integrand in a similar manner to the Korobov transforms. One potentially negative feature of the Sidi transforms is that several trigonometric functions need to be computed for each sample of the integrand. This can increase the cost (in terms of machine operations) considerably, especially for relatively simple integrands.

The baker's transformation~\cite{baker_trafo} (also called ``tent transformation''), given by
\begin{align}
\phi(u) = 1- \big| 2u-1\big|=&\left\{
\begin{tabular}{ll}
$2u$ & $\mbox{ if } u\leq \frac{1}{2},$\\
$2-2u$ & $\mbox{ if } u> \frac{1}{2},$
\end{tabular}
\right.                   \label{eq:baker}           
\end{align}
can be applied to achieve close to $\mathcal{O}(n^{-2})$ convergence for non-periodic
integrands. The transform periodizes the integrand by mirroring rather than forcing it to a particular value on the integration boundary. Naively the fact that the transform is discontinuous might lead us to expect a poor asymptotic scaling (due to the fact that the transform is not smooth). However, an analysis based on considering the transform as a modification of the lattice rather than of the integrand allows the convergence of $\mathcal{O}(n^{-2})$ to be proven. In a moderate number of dimensions ($d \gtrsim 9$) the baker's transform typically does not increase the variance of the integrand as much as the Korobov and Sidi transforms. Therefore, although it has a slower convergence rate, the baker's transform can still prove useful.

A critically important point to consider when choosing a periodization strategy is the number of dimensions in which the integration will be performed. In particular,
applying a periodizing transform can increase the variance of the integrand exponentially with its dimension $d$.
Although it is possible to construct rank-1 lattice rules whose worst case error is independent of $d$ (or depends at most polynomially on $d$), increasing the variance of the integrand
can spoil the convergence of the quasi-Monte Carlo integration~\cite{Kuo2007}.
For integrands in a relatively low number of dimensions ($d\lesssim 8$) the increase in variance caused by higher weight
($r\gtrsim 3$) periodizing transforms can be counteracted by the improved smoothness of the integrand which leads to an improved asymptotic scaling behaviour with the number
of lattice points $n$.

\subsection{Variance reduction}
\label{sec:mathIS}
By applying a variable transformation $y=p(x)$ to a one-dimensional integral
\begin{equation}
  \label{eq:ISintegral}
  I=\int_0^1 \mathrm{d}y\;f(y) = \int_0^1 \mathrm{d}x\;p'(x)\,f(p(x)),
\end{equation}
the integration becomes trivial if $p'(x)\propto f(p(x))^{-1}$. While it is usually not possible to find a transformation fulfilling this condition exactly, it is possible to find approximations to it. This leads to an integrand with reduced variance, which can significantly improve the convergence of the integration when using numerical integration techniques. 
A well known method to apply variance reduction to multi-dimensional integrals is the \vegas{} algorithm\,\cite{Lepage:1977sw}, where the above procedure is applied to each integration variable separately, with the remaining variables integrated out. In the algorithm of Ref.~\cite{Lepage:1977sw}, for each integration variable $x$, the transformation $p(x)$ is constructed as a strictly increasing, piecewise linear function, such that $p'(x)$ resembles the shape of $|f(p(x))|^{-1}$. However, this procedure leads to discontinuities in $p'(x)$, which spoil the smoothness of the integrand and thus the scaling of the numerical integration when directly applying this algorithm in combination with \qmc{} integration. Instead, we use the ansatz
\begin{equation}
  \label{eq:ISfit}
  p(x) = a_2 \cdot x\,\frac{a_0-1}{a_0-x} + a_3 \cdot x\,\frac{a_1-1}{a_1-x} + a_4\cdot x + a_5\cdot x^2 + \left(1-\sum_{i=2}^5 a_i\right)\cdot x^3
\end{equation}
to parametrize the variance reducing transformation. The parameters $a_i$ are obtained via a fit to the inverse of the cumulative distribution function (CDF),
\begin{equation}
  \mathrm{CDF}_f(x) = \int_0^x \mathrm{d}y\;|f(y)|\, \Big/ \int_0^1 \mathrm{d}y\;|f(y)|.
\end{equation}
The ansatz in Eq.~(\ref{eq:ISfit}) is chosen such that $p(0)=0$ and $p(1)=1$. The parameters $a_2$ and $a_3$ are required to be positive, and $a_0\in [1.001,\infty)$, $a_1\in(-\infty,-0.001]$ such that no singularities are introduced within the domain of integration by the transforms.
The parameters are optimized by
sampling the integrand with a lattice of given size to numerically obtain an estimate of the CDF for each integration parameter and applying a non-linear least-squares fit using the routines implemented in the GNU Scientific Library~\cite{Gough:2009:GSL:1538674}.

We find that the ansatz in Eq.~(\ref{eq:ISfit})  works well for typical functions obtained by sector decomposition. 
While this ansatz  in principle can be applied to other integrals as
well, we expect that for other functions it can be beneficial to modify it to improve the fit of the CDF of the corresponding integrand.

\section{Stand-alone usage of the integrator library}
\label{sec:standalone}
\subsection{Installation}

If you wish to use the integrator with your own code rather than within \pysecdec{}, then it is available as a \cppeleven{} single-header header-only library at \href{https://github.com/mppmu/qmc}{https://github.com/mppmu/qmc}. Download the header and include it in your project.
Since the \qmc{} is a header only \cpp{} template library it does not need to be separately configured and built. 

In order to build the header as part of your project you will need:
\begin{itemize}
\item A \cppeleven{}  compatible \cpp{} compiler.
\item The GNU Scientific Library~(GSL), version 2.5 or greater.
\item (Optional GPU support)  A \cuda{} compatible compiler (typically \texttt{nvcc}).
\item (Optional GPU support) \cuda{} compatible hardware with Compute Capability 3.0 or greater.
\end{itemize}
Simply include it in your project, ensure that it can be found by your compiler (using compiler include path specifiers if necessary) and then build your project, linking against the GSL.

\subsection{Minimal example}

In this section we provide examples of the usage of the integrator as a stand-alone package. The usage within the \cpp{} interface to \pysecdec{} is similar while the usage via the python interface to \pysecdec{} differs significantly.
Both uses within \pysecdec{} are described in Section \ref{subsec:usagepysecdec}.

The code of a minimal program demonstrating the usage of the integrator is shown in Fig.~\ref{fig:cppminimal}. In this example, the 3-dimensional function $f(x_0,x_1,x_2) = x_0 x_1 x_2$ is integrated using the default settings of the \qmc{}. Assuming the code is in a file named \texttt{minimal.cpp} and the \qmc{} header can be found by the compiler, the program can be compiled without GPU support using the command:
\begin{lstlisting}[language=bash,numbers=none,basicstyle=\scriptsize,breaklines=false]
c++ -std=c++11 minimal.cpp -o minimal.out -lgsl -lgslcblas
\end{lstlisting}
or with GPU support using the command:
\begin{lstlisting}[language=bash,numbers=none,basicstyle=\scriptsize,breaklines=false]
nvcc -arch=<arch> -std=c++11 -x cu -Xptxas -O0 -Xptxas 
--disable-optimizer-constants minimal.cpp -o minimal.out -lgsl -lgslcblas
\end{lstlisting}
where \texttt{<arch>} is the architecture of the target GPU or \texttt{compute\_30} for just-in-time compilation (see the Nvidia \texttt{nvcc} manual for more details). The compile flag \texttt{-x cu} explicitly specifies the language of the input files as \cuda{}, rather than letting the compiler choose a default based on the file name suffix. The compile flag \texttt{-Xptxas -O0} disables optimisation of the code by the PTX assembler, as of \cuda{} 9.2 we found rare cases where code optimisation led to wrong results. The flag \texttt{-Xptxas --disable-optimizer-constants} disables the use of the optimizer constant bank which can be exhausted for large integrands, it is not strictly necessary to pass this flag for simple examples.

In Fig.~\ref{fig:cppminimal}, on lines $4-13$, a functor \texttt{my\_functor}, containing the function to be integrated is defined and instantiated.
On line $19$ the \qmc{} integrator is instantiated with a Korobov transform of weight 3. The \texttt{MAXVAR} variable controls the maximum number of integration variables over which a particular instance of the \qmc{} integrator can integrate, it should be set to a value equal to or larger than the maximum number of integration parameters present in any functor that will be passed to the instance of the \qmc{} integrator.
On line $20$ the functor instance is passed to the \texttt{integrate} function of the integrator, this will trigger the numerical integration.
The integrator returns a \texttt{result} struct containing the integral and its uncertainty, which are printed on lines $21-22$. 
The \cuda{} function execution space specifiers \texttt{\_\_host\_\_} and \texttt{\_\_device\_\_} on line $7$ are present only when compiling with GPU support. This is controlled by the presence on line $6$ of the \texttt{\_\_CUDACC\_\_} macro which is automatically defined by the compiler during \cuda{} compilation.

\begin{figure}
\begin{center}
\begin{lstlisting}
#include <iostream>
#include "qmc.hpp"

struct my_functor_t {
    const unsigned long long int number_of_integration_variables = 3;
#ifdef __CUDACC__
    __host__ __device__
#endif
    double operator()(double* x) const
    {
        return x[0]*x[1]*x[2];
    }
} my_functor;

int main() {

    const unsigned int MAXVAR = 3;

    integrators::Qmc<double,double,MAXVAR,integrators::transforms::Korobov<3>::type> integrator;
    integrators::result<double> result = integrator.integrate(my_functor);
    std::cout << "integral = " << result.integral << std::endl;
    std::cout << "error = " << result.error << std::endl;

    return 0;
}
\end{lstlisting}
\end{center}
\caption{A minimal example of the use of the \qmc{} integrator.}
\label{fig:cppminimal}
\end{figure}

\subsection{Usage}

We envisage two typical use case scenarios for the \qmc{} library:
\begin{enumerate}
\item The user knows relatively little about the integrand but wishes to know the result with a specific relative and/or absolute accuracy.
\item The user has a reasonable idea how the \qmc{} performs on their integrand and wishes to obtain a result as quickly as possible.
\end{enumerate}
We discuss these use cases in turn. A description of all public fields and member functions is given in Appendix~\ref{sec:appendix:api}.

\subsubsection{Case 1}

In order to evaluate an integral to a specific relative and/or absolute accuracy without significant human input, the following \qmc{} integrator member variables are relevant: \texttt{epsrel}, \texttt{epsabs}, \texttt{maxeval} along with the member function \texttt{integrate}.

Firstly, the \qmc{} must be initialised with a suitable integral transform and the user must decide whether to use the variance reduction methods described in Section~\ref{sec:mathIS}. If nothing is known about the periodicity and variance of the integrand, we would typically recommend using a Korobov weight 3 transform if the integrand lives in less than 9 dimensions (otherwise the baker transform may be more suitable) and no variance reduction. If the \qmc{} does not produce even a rough estimate of the integral ($\sim 20\%$ error) with a moderate lattice size then the variance reduction procedure may prove useful and the fit function should be specified as shown in Fig.~\ref{fig:case1}.

The user can then set the \texttt{epsrel} and \texttt{epsabs} fields to the desired accuracy. In addition, the  parameter \texttt{maxeval} ensures that the integration terminates in a reasonable time, even if the desired accuracy cannot be reached. 
The integration terminates once any of the three conditions is met.
What constitutes a suitable value of \texttt{maxeval} depends on the complexity of the integrand (in terms of floating point operations), the hardware available for computing the integral and the time the user is willing to wait for a result.

Finally, the \texttt{integrate} function can be called on the input function. In Fig.~\ref{fig:case1} we display the above steps in code (for a $5$-dimensional real integrand named \texttt{my\_integrand} ).

\begin{figure}
\begin{center}
\begin{lstlisting}
integrators::Qmc<double,double,5,integrators::transforms::Korobov<3>::type
                 ,integrators::fitfunctions::PolySingular::type // optional
                > integrator;
integrator.epsrel = 1e-5; // requested relative accuracy
integrator.epsabs = 1e-20; // requested absolute accuracy
integrator.maxeval = 10000000; // maximum number of function evaluations
integrators::result<double> result = integrator.integrate(my_integrand);
\end{lstlisting}
\end{center}
\caption{Case 1 usage example of the \qmc{} integrator.}
\label{fig:case1}
\end{figure}

If a fit function has been provided, the \qmc{} library will evaluate \texttt{evaluateminn} lattice points and use them as input to the fitting and variance reduction procedure as described in Section~\ref{sec:mathIS}. The \qmc{} library will then apply the selected periodizing transform to the fitted function. If no fit function has been provided the \qmc{} will apply the periodizing transform to the input function and proceed to the next step directly. 

In the next step, a total of \texttt{minm} randomly shifted copies of the smallest possible lattice greater than \texttt{minn} in size will be sampled and used to estimate the integration error. If the required error goal has not been reached the result will be discarded and a larger lattice will be selected and computed. This procedure will be repeated as necessary until the desired error goal is reached or \texttt{maxeval} function evaluations have been performed; at which point the integration will terminate and the last result obtained will be returned. 

If, during the iteration, the \qmc{} requires a lattice larger than can be produced with the available generating vectors it will instead select the largest lattice and attempt to reduce the integration error by adding random shifts. In this case the \qmc{} will achieve only Monte Carlo $\mathcal{O}(n^{-1/2})$ scaling. If an acceptable result can not be achieved with Monte Carlo scaling then the user is advised to compute and supply additional (larger) generating vectors as described in Section~\ref{sec:appendix:generatingVectors}.

Note that, unlike some other integration algorithms, the results from all but the last iteration have no effect on the final result. It is therefore always more efficient to directly evaluate a lattice that gives an acceptable integration error rather than asking the library to try to find a suitable lattice size by iterating.

\subsubsection{Case 2}
\label{sec:use_case_2}

If the user has a reasonable idea how the \qmc{} performs on their integrand, for example by studying similar integrands or evaluating their integrand with a small lattice, then a result can most quickly be obtained by setting the parameters \texttt{minn} and calling the member function \texttt{integrate}. In order to ignore the default error goals \texttt{epsrel} and \texttt{epsabs}, the parameter \texttt{maxeval} should be set to $1$. In Fig.~\ref{fig:case2} we display the above steps in code (for a $5$-dimensional real integrand named \texttt{my\_integrand}).

\begin{figure}
\begin{center}
\begin{lstlisting}
integrators::Qmc<double,double,5,integrators::transforms::Korobov<3>::type> integrator;
integrator.minn = 10000; // suitable lattice size
integrator.maxeval = 1; // do not evaluate larger lattices to satisfy default epsrel and epsabs
integrators::result<double> result = integrator.integrate(my_integrand);
\end{lstlisting}
\end{center}
\caption{Case 2 usage example of the \qmc{} integrator.}
\label{fig:case2}
\end{figure}

The \qmc{} library will evaluate \texttt{minm} randomly shifted copies of the smallest possible lattice with at least \texttt{minn} points and return the result. If this result is not satisfactory then the user can increase \texttt{minn} and retry the integration. In order to estimate what lattice size is suitable it is sometimes useful to investigate the scaling behaviour of the integrand by evaluating several different lattices. Note that, as can be seen in Fig.~\ref{fig:scalingplot}, the scaling of the \qmc{} is quite `noisy` in the sense that very similarly sized lattices can produce estimates of the integral with errors that differ by an order of magnitude or more. This behaviour can hinder straightforward attempts to estimate the scaling behaviour of an integrand.

\subsubsection{Usage on GPUs }

In order to use the \qmc{} library on \cuda{} enabled GPUs the user must ensure that their integrand functor can be evaluated on the chosen device. This usually entails taking the following steps:
\begin{itemize}
\item Ensuring that the \cpp{} language features used in the integrand function are supported by the relevant \cuda{} device.
\item Designing the integrand function so that it does not need to access data that will be stored only in host memory.
\item Marking the call operator of the integrand functor \texttt{\_\_host\_\_\ \_\_device\_\_}, as shown in the examples above.
\end{itemize}

For the purpose of monitoring GPU usage and debugging we have found the following tools provided by Nvidia, and distributed with the \cuda{} toolkit, to be useful:
\begin{itemize}
\item \texttt{nvidia-smi}, a \texttt{top} like management and monitoring utility for Nvidia GPU devices.
\item \texttt{cuda-memcheck}, a functional correctness checking suite.
\end{itemize}

In most cases the usage of GPUs within the \qmc{} is straightforward, however, the attentive user may notice that the program behaves in a slightly different manner than when using only CPUs. Let us discuss some of the most prominent features of \cuda{} devices which can affect the usage of the \qmc{} library.

The Nvidia kernel mode driver must be running and connected to the GPU device before any user interaction with that device can take place. If the kernel mode driver is not already running and connected to the target GPU the invocation of a program that interacts with the GPU will cause the driver to load and initialize the GPU. This will incur a start up cost of 1-3 seconds per GPU. For short running integration jobs this cost can be a significant fraction of the integration time. On Windows, the kernel mode driver is loaded at start up and kept loaded until shut down, however, by default the time-out detection and recovery (TDR) feature will cause driver reload and should be disabled (we refer to the latest Nvidia documentation). Similarly, under Linux, if an \texttt{X}-like process is run from start up to shut down it will usually initialize and keep alive the kernel mode driver. However, if no long-lived \texttt{X}-like client is kept running (for example in many HPC environments) the kernel mode driver will initialize and de-initialize the target GPU each time a GPU application starts and stops. As of \cuda{} 9.2 the Nvidia recommended way to circumvent the delay due to starting and stopping the kernel mode driver is to run the Nvidia Persistence Daemon (we refer to the latest Nvidia documentation).

When compiling with \texttt{nvcc} we strongly recommend to look up and enter the \emph{real} architecture of the graphics card in use, e.g. {\tt -arch=sm\_70}. If a \emph{virtual} architecture is specified, the device code is just-in-time compiled for the \emph{real} architecture on a single core, which may become the dominant fraction of the runtime. For initial tests, the \emph{virtual} architecture {\tt compute\_30}, which is the oldest supported in \cuda{} version 9.2, should be compatible with most GPUs that are currently in use. For more information we refer to the {\tt nvcc} manual.

\subsection{Design and Implementation}
\label{sec:implementation}

In order to numerically integrate a single function, the \qmc{} integrator library can concurrently utilise multiple multi-threaded CPUs as well as multiple \cuda{} hardware accelerators, provided they all belong to a single system. To achieve reasonable performance on heterogeneous systems a receiver-initiated central work queue load balancing algorithm is utilised. 

The load balancing algorithm consists of the following steps:
\begin{itemize}
\item A central work queue is initialised by the main thread. 
\item The main thread then spawns \texttt{cputhreads} worker threads if \texttt{-1} is listed in  \texttt{devices} and additionally one worker thread per GPU listed in \texttt{devices}.
\item Each worker requests work from the work queue and when the work is completed continues to request work until the queue is cleared, at which point the workers terminate. 
\end{itemize}
The disadvantage of this algorithm is that the central work queue must be atomically locked to ensure work is not repeated. This can impact performance significantly when a quick-to-evaluate integrand is computed using a large number of cores and/or \cuda{} devices. 
The advantage of this design is that even if the performance of the workers differs vastly (for example a worker computing on a single CPU core compared to a worker distributing work to a powerful GPU) the workload is reasonably balanced provided that the work packages are not so large as to leave a poorly performing worker with so much work that it finishes significantly later than all others. 

In order to utilise massively parallel \cuda{} hardware, threads assigned to provide work to an accelerator will request significantly more work from the queue per access than workers assigned to a single CPU core. The amount of work (number of ``work packages'') requested at once by a worker assigned to a \cuda{} device is controlled by the product of \texttt{cudablocks} and \texttt{cudathreadsperblock}, while workers assigned to a CPU core always request only a single ``work package''.

At the time of release the default values of parameters affecting the load balancing are usually a reasonable choice for most feasible integrands and existing hardware. Naturally, as the state of the art advances and computer hardware evolves we may alter these default values in future \qmc{} releases.

\section{Usage of the integrator library within \pysecdec}
\label{sec:pysecdec}
Here we describe briefly the installation and usage of \pysecdec{},
focusing on the usage of the \qmc{} integrator.
For more details we refer to the manual \url{https://secdec.readthedocs.io}
and to the examples distributed with \pysecdec{}.

\subsection{Installation}

Before installing \pysecdec{}, make sure that recent versions of numpy (\url{http://www.numpy.org/}) and sympy (\url{http://www.sympy.org/}) are installed.
The \pysecdec{} program (which includes the \qmc{} integrator library) can be downloaded from \url{https://github.com/mppmu/secdec/releases}.
To install \pysecdec{}, perform the following steps

{\tt
tar -xf pySecDec-<version>.tar.gz \\
cd pySecDec-<version> \\
make \\
<copy the highlighted output lines into your .bashrc>
}

The \texttt{make} command will automatically build further dependencies
in addition to \pysecdec{} itself.
Further notes on the installation procedure are summarized in the online documentation \url{https://secdec.readthedocs.io}.
To get started, we recommend to read the section ``getting started" in the online documentation.

\subsection{Usage}
\label{subsec:usagepysecdec}

Depending on the availability, it is possible to use the program with CPUs, GPUs 
or a combination of both. 

\subsubsection{Using CPUs only}
\label{subsubsec:cpus}
The basic steps can be summarized as follows:
\begin{enumerate}
\item Write or edit a \python{} script to define the integral, 
the replacement rules for the kinematic invariants,
the requested order in the regulator and some other options, see e.g. 
the example {\tt examples/easy/generate\_easy.py}.
\item Run the script 
{\tt generate\_easy.py}
using \python{}. This will generate a subdirectory according 
to the {\tt name} specified in the script.
\item Type {\tt make -C <name>}, where {\tt <name>} is your chosen name. This will create the \cpp{} libraries.
\item Write or edit a \python{} script to perform the numerical integration using the
  \python{} interface, see e.g. {\tt examples/easy/integrate\_easy.py}.
  Make sure that the \qmc{} integrator is chosen in that file.
\end{enumerate}

\subsubsection{Using GPUs and CPUs}

When using GPUs, steps (1), (2) and (4) of the previous section~\ref{subsubsec:cpus} are the same. The only difference is in the compilation of the sector files
\begin{enumerate}
\item[(3)] Type {\tt CXX=nvcc SECDEC\_WITH\_CUDA=<arch> make -C <name>}, where {\tt <name>} is your chosen name and {\tt <arch>}
is the argument forwarded to {\tt nvcc} as {\tt -arch=<arch>}. This will create the \cpp{} libraries.
\end{enumerate}

The compute capability {\tt <arch>} is specific to each graphics card. The parameter {\tt <arch>} can either be a suitable \emph{virtual} architecture or a \emph{real} architecture.
We strongly recommend to look up and enter the \emph{real} architecture of the graphics card in use, e.g. {\tt sm\_70}. If a \emph{virtual} architecture is specified, the device
code is just-in-time compiled for the \emph{real} architecture on a single core, which may become the dominant fraction of the runtime. For first tests however, the \emph{virtual} architecture
{\tt compute\_30}, which is the oldest supported in \cuda{} version 9.2, should be compatible with most GPUs that are currently in use.
For more information refer to the {\tt nvcc} manual.

\subsection{Examples}
\label{sec:examplespysd}
All the examples described below can be found in the folder {\tt examples} of the \pysecdec{} distribution. 
A comparison of the timings for the examples can be found in Table~\ref{tab:timings_pysd}.
The settings for the examples are default settings unless specified otherwise. The setting {\tt maxeval=1}
ensures the evaluation of the integrand with a fixed number of sampling points as described in Section~\ref{sec:use_case_2}.

\subsubsection{Basic usage}

The basic usage of \pysecdec{} is illustrated in the example \texttt{easy}. The slightly modified \texttt{easy\_cuda} example
shows how to compile and run the easy example on all available GPUs using either the \python{} or the \cpp{} interface.

\begin{figure}[htb]
    \begin{subfigure}[b]{0.5\textwidth}
\begin{lstlisting}[
            language=python,
            xleftmargin=7ex,
            keywordstyle=\color{cyan}\ttfamily,
            commentstyle=\color{blue}\ttfamily
        ]
from pySecDec import make_package

make_package(

name = 'easy',
integration_variables = ['x','y'],
regulators = ['eps'],

requested_orders = [0],
polynomials_to_decompose = ['(x+y)^(-2+eps)'],

)
\end{lstlisting}
        \caption{\tt generate\_easy.py}
    \end{subfigure}
        \begin{subfigure}[b]{0.5\textwidth}
\begin{lstlisting}[
            language=python,
            xleftmargin=7ex,
            keywordstyle=\color{cyan}\ttfamily,
            commentstyle=\color{blue}\ttfamily
        ]
from pySecDec.integral_interface import IntegralLibrary
from math import log

# load c++ library
easy = IntegralLibrary('easy/easy_pylink.so')

# choose Qmc integrator
# automatically uses all avaliable GPUs
easy.use_Qmc(transform='korobov3')

# integrate
_, _, result = easy()

# print result
print('Numerical Result:' + result)
print('Analytic Result:' + ' + (%.15g)*eps^-1 + (%.15g) + O(eps)' % (1.0,1.0-log(2.0)))
\end{lstlisting}
        \caption{\tt integrate\_easy.py}
    \end{subfigure}
    \caption{\pysecdec{} input for a simple integral.}
    \label{fig:easy_example}
\end{figure}

The generate file {\tt generate\_easy.py} shown in Fig.~\ref{fig:easy_example} is identical for both examples, \texttt{easy} and \texttt{easy\_cuda}.
The integrate file {\tt integrate\_easy.py} differs by the optional lines that select the \qmc{} integrator. Choosing the \qmc{}
integrator as shown in Fig.~\ref{fig:easy_example} will make \pysecdec{} use all CPU cores as well as all available GPUs.

In order to use GPUs, the code should be compiled with Nvidia's {\tt nvcc} compiler. It is also possible to use non-\cuda{} compilers, though this will disable GPU support.

The commands to run the examples are\\
(a) using CPUs and GPUs:\\
{\tt 
python generate\_easy.py\\
CXX=nvcc SECDEC\_WITH\_CUDA=compute\_30 make -C easy\\
python integrate\_easy.py
}

\noindent or (b) using CPUs only:\\
{\tt 
python generate\_easy.py\\
make -C easy\\
python integrate\_easy.py\\
}

How to use the \qmc{} integrator with GPU support via the \cpp{} interface is shown in the example \texttt{easy\_cuda}.
Note that above we have set the compute capability to {\tt compute\_30}. Please read the remarks about the compute capability
in the previous section before running more complicated examples on the GPU.

\subsubsection{3-mass banana graph}

\begin{figure}[htb]
  \centering
  \includegraphics[width=0.4\textwidth]{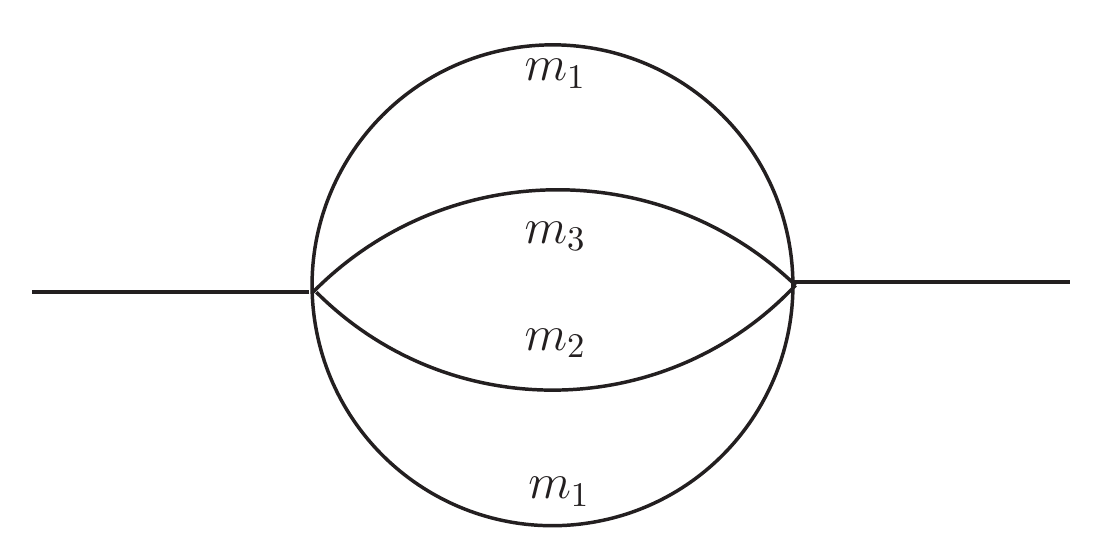}
  \caption{A 3-loop 2-point function with 3 different masses.}
  \label{fig:banana3m}
\end{figure} 

The example \texttt{banana\_3mass} calculates a three-loop two-point 
integral with three different internal masses, see 
Fig.~\ref{fig:banana3m}. If the three masses are different, the 
analytic result cannot be expressed anymore by products of complete
elliptic integrals~\cite{Primo:2017ipr}, and therefore is an example
of a ``hyperelliptic'' integral.
The purpose of this example is to show that hyperelliptic 
integrals can be evaluated as fast as other integrals which are 
more accessible analytically. 

\medskip

The result for the non-Euclidean point $s=20.0$,
$m_1^2=1.0$, $m_2^2=1.3$, $m_3^2=0.7$, computed with the
\qmc{} and the settings {\tt minn=1000000}, {\tt maxeval=1}, {\tt transform=`korobov2'} reads
\begin{align}
I&=
(\, 1.97000000000000264 \pm 9.85\cdot 10^{-15}  + \text{i}\,(1.84\cdot 10^{-15} \pm  1.16\cdot 10^{-15} )) \cdot\eps^{-3}\nn\\
&+(\,-5.9281676367925620 \pm  6.52\cdot 10^{-14} + \text{i}\,(1.07\cdot 10^{-13} \pm 2.63\cdot 10^{-14} ))\cdot\eps^{-2}\nn\\
& +(\,  9.86757086818429 \pm  1.64\cdot 10^{-12}- \text{i}\,(2.54\cdot 10^{-11}\pm 9.83\cdot 10^{-12} ))\cdot\eps^{-1}\nn\\
&  \quad -89.066074732329 \pm  8.25\cdot 10^{-10}  + \text{i}\,(8.10892634289 \pm 2.37\cdot 10^{-9} )\nn\\
  &+ \mathcal{O}(\eps)\;.
\end{align}
The imaginary parts of the pole coefficients are numerically zero, the
accuracy of this zero being limited by the fact that we are operating
close to machine precision.

\subsubsection{Non-planar 4-point function with massive propagators and massive legs of different mass}

\begin{figure}[htb]
  \centering
  \includegraphics[width=0.35\textwidth]{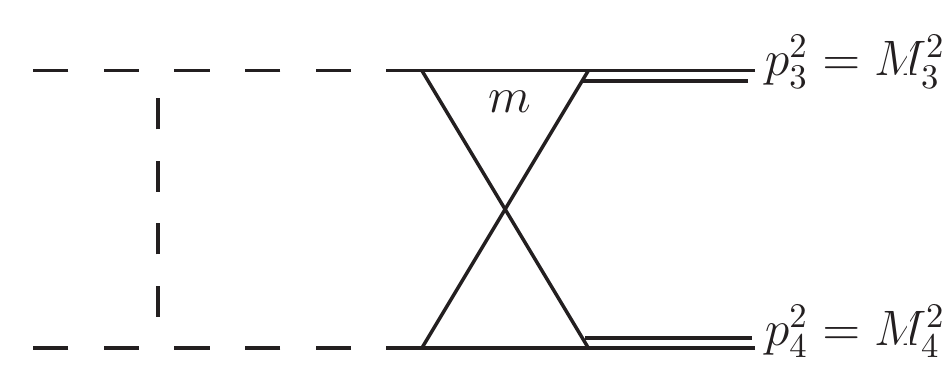}
  \caption{A non-planar 2-loop 4-point function with a massive loop and two massive legs with different masses.}
  \label{fig:HZ}
\end{figure} 

The example \texttt{HZ2L\_nonplanar} calculates a non-planar four-point 
two-loop integral in the physical region, where one loop is fully 
massive, and two of the external legs are massive/off-shell with two different masses, see Fig.~\ref{fig:HZ}. 
The commands to run this example are analogous to the ones given above.

\medskip

The result for the point $s=200$, $t=-23$, $m^2=9,M_3^2=1.56, M_4^2=0.81$, 
obtained with the \qmc{} integrator using the settings {\tt  minn=10**8}, {\tt maxeval=1}, {\tt transform=`korobov3'}  reads
\begin{align}
&(\,3.4401552304457233\cdot 10^{-6} \pm 1.73\cdot 10^{-20}
  - \text{i}\,(8.9\cdot 10^{-23} \pm 2.26\cdot 10^{-21} ))\cdot\eps^{-2}\nn\\
+&(\, -0.00003316795824 \pm  1.16\cdot 10^{-12}
  - \text{i}\,(8.53692099\cdot 10^{-6} \pm 1.01\cdot 10^{-12} ))\cdot\eps^{-1}\nn\\
+&\quad 0.000159345747 \pm  4.12\cdot 10^{-10} 
+ \text{i}\,(0.000021017686 \pm 3.89\cdot 10^{-10} )\nn\\
  +& \mathcal{O}(\eps)\;.
\end{align}

\subsubsection{Pentabox}

\begin{figure}[htb]
  \centering
  \includegraphics[width=0.35\textwidth]{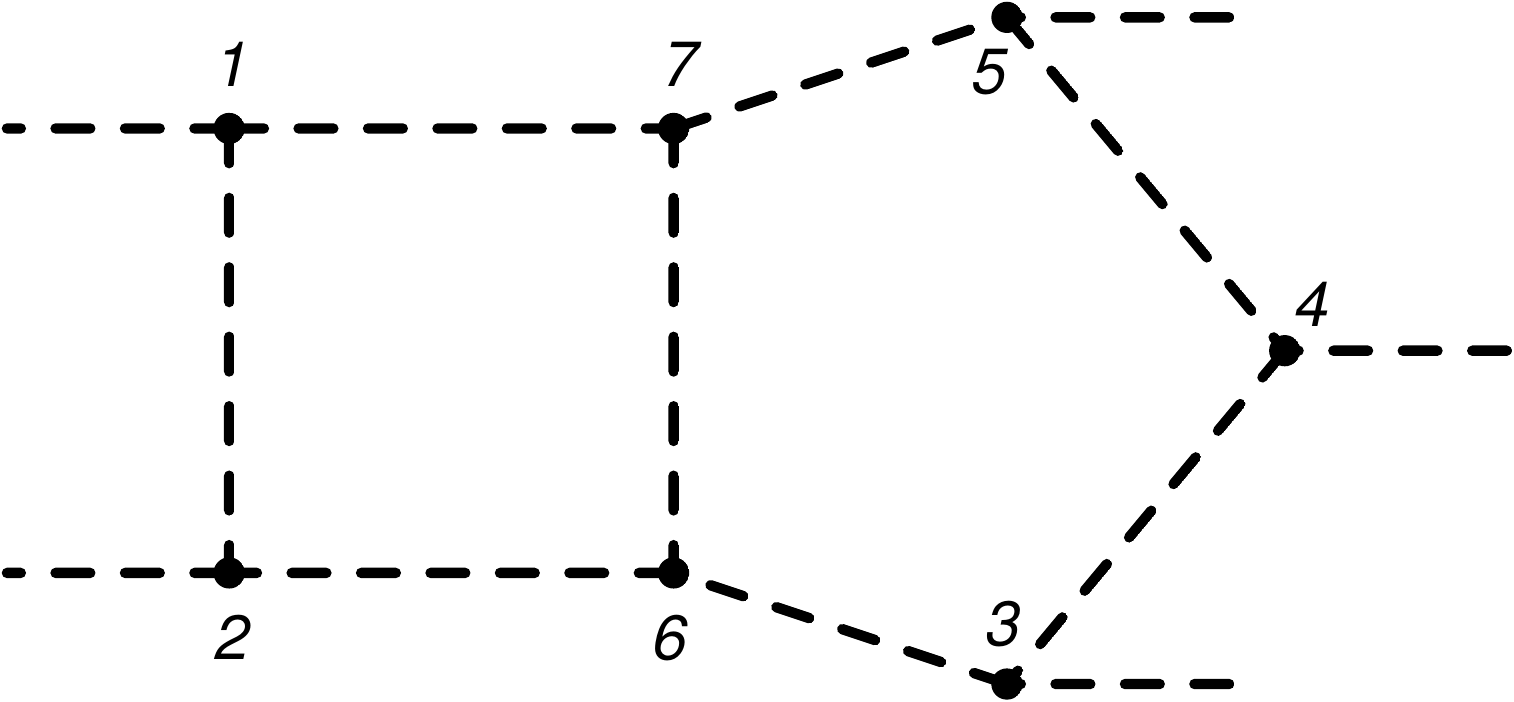}
  \caption{A 2-loop pentabox integral in $d=6-2\eps$.}
  \label{fig:pentabox}
\end{figure} 
The example \texttt{pentabox\_fin} calculates a fully massless 
two-loop five-point function in the physical region with 
$d=6-2\eps$, see Fig.~\ref{fig:pentabox}.

The pentabox is a master integral occurring in the calculation of $2 \rightarrow 3$ scattering at two loops. 
In $4-2\eps$ dimensions, sector decomposition produces poles of order $\eps^{-5}$ at intermediate stages. 
The $6-2\eps$ dimensional version we investigate here is finite and therefore a more suitable master integral for numerical evaluation.
This is an example where the use of a fit function considerably improves the convergence.

\medskip

The result for the non-Euclidean point $s_{12}=5$, $s_{23}=-4$, 
$s_{34}=2$, $s_{45}=-6$ and $s_{51}=3$, obtained with the \qmc{} integrator
using the settings 
{\tt minn=10**8}, {\tt maxeval=1}, {\tt transform=`korobov3'}, {\tt fitfunction=`polysingular'} 
reads
\begin{align}
P=& -0.0198236478 \pm 2.02\cdot10^{-8} - \text{i}\,(\, 0.0341514614 \pm 1.59\cdot10^{-8}\, )
 + \mathcal{O}(\eps)\,.
\end{align}

\subsubsection{Elliptic 2-loop integral}

\begin{figure}[htb]
  \centering
  \includegraphics[width=0.3\textwidth]{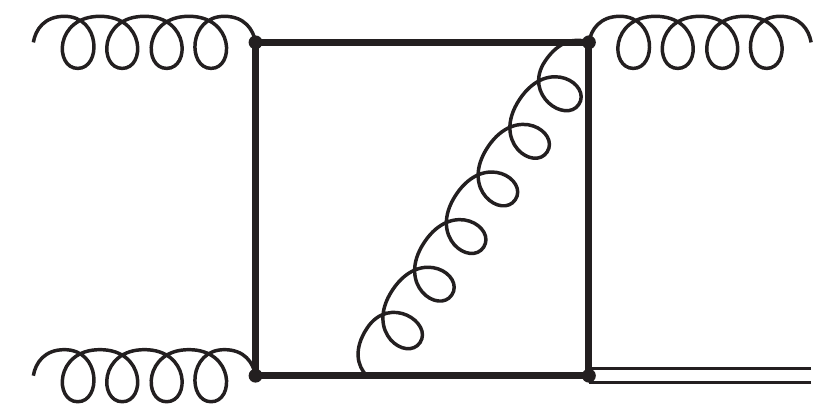}
  \caption{A 2-loop 2-point integral appearing at NLO in Higgs plus jet production.}
  \label{fig:box2L_ell}
\end{figure} 

The example \texttt{elliptic2L\_physical} calculates a planar 
two-loop four-point 
function with one off-shell leg and a massive loop in the physical 
region, see Fig.~\ref{fig:box2L_ell}. This diagram enters the NLO
corrections to Higgs$+$jet 
production and contains elliptic structures. The analytical result 
in the Euclidean region is given in Ref.~\cite{Bonciani:2016qxi}.
While a numerical result for this integral already has been given in
Ref.~\cite{Borowka:2017idc}, the purpose of this example is to
demonstrate that the number of correct digits which can be obtained
using the \qmc{} integrator cannot be reached in a reasonable amount of time using 
Monte Carlo integration.

\medskip

The result for the non-Euclidean point $s=90$, $t=-2.5$, $p_4^2=1.6,
m^2=1$ using \vegas{} reads
\begin{align} 
f_{66}^A\cdot \left(\frac{-s}{m^2}\right) = -0.044289 \pm 2.5\cdot10^{-5} + \text{i}\, (\,0.016068 \pm 2.7\cdot10^{-5}\,) + \mathcal{O}(\eps)\,.
\end{align}
Using the \qmc{} integrator with {\tt minn=2147483647}, {\tt maxeval=1}, \\ {\tt transform=`korobov1'}, {\tt fitfunction=`polysingular'}:
\begin{align}
\begin{split}
f_{66}^A\cdot \left(\frac{-s}{m^2}\right) =
  &  -0.04429245890863\pm  1.82\cdot 10^{-13} \nn\\+&\; \text{i} \,(\, 0.01607147782349   \pm 1.69\cdot 10^{-13}\,) + \mathcal{O}(\eps)\,.
\end{split}
\end{align}


\subsubsection{Hyperelliptic 2-loop integral}

\begin{figure}[htb]
  \centering
  \includegraphics[width=0.3\textwidth]{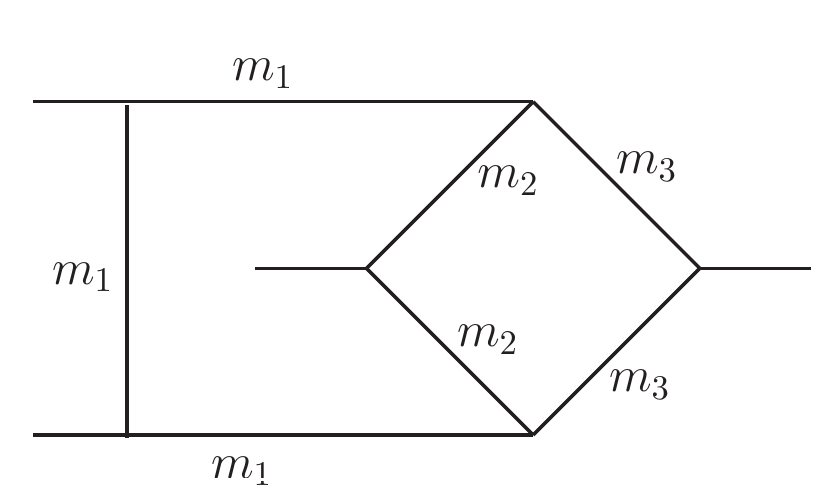}
  \caption{A 2-loop integral leading to hyperelliptic functions~\cite{Georgoudis:2015hca}.}
  \label{fig:hyperelliptic}
\end{figure} 

The example \texttt{hyperelliptic} calculates a non-planar two-loop 
four-point function with three different masses and all propagators 
massive in the physical region, see Fig.~\ref{fig:hyperelliptic}. 
This integral is special since it is extremely hard to compute analytically, 
but is easily accessible numerically. 
\medskip

The result for the non-Euclidean point $s=10$, $t=-0.75$, $m_1^2=1$, $m_2^2=1.3$,
$m_3^2=0.7$, computed with the \qmc{} integrator using {\tt minn=10**8}, {\tt maxeval=1}, {\tt transform=`korobov3'}, {\tt fitfunction=`polysingular'}
reads
\begin{align}
I = & -0.009449626 \pm  1.54\cdot 10^{-7} + \text{i} \,(\,0.019368308 \pm 1.60\cdot 10^{-7}\,)
 + \mathcal{O}(\eps)\,.
\end{align}

\subsubsection{4-loop form factor example}

\begin{figure}[htb]
  \centering
  \includegraphics[width=0.3\textwidth]{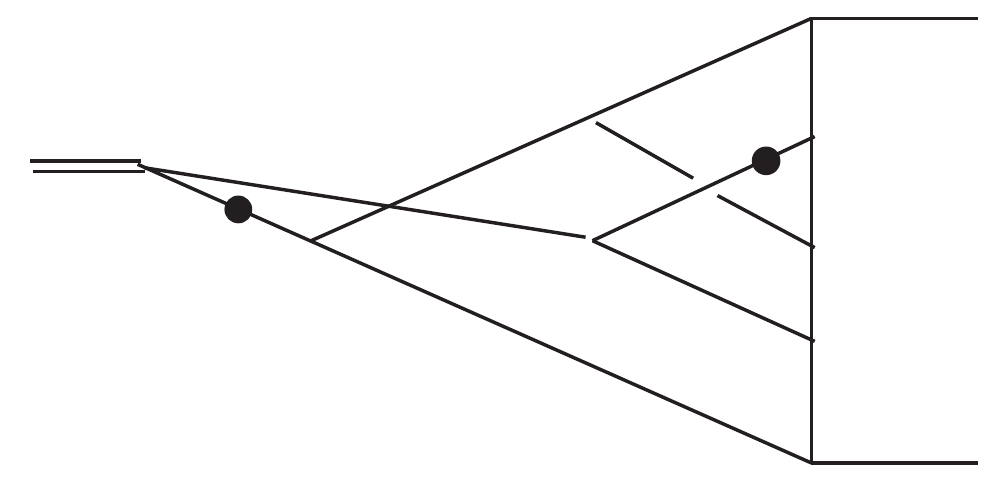}
  \caption{A 4-loop massless form factor integral~\cite{vonManteuffel:2015gxa}.}
  \label{fig:4LFF}
\end{figure} 

The example \texttt{formfactor4L} calculates a four-loop three-point integral 
in $d=6-2\eps$, see Fig.~\ref{fig:4LFF}. Its analytic result is given in Eq.~(7.1) of 
Ref.~\cite{vonManteuffel:2015gxa}. This example demonstrates the 
power of \pysecdec{} to perform an efficient sector 
decomposition, even for integrals with many loops and internal 
propagators. Furthermore, it is a prime example to show how the 
\qmc{} algorithm works for a larger number of integration dimensions (in this case 11 dimensions).

Since the integral has only one scale, the latter can be factorized. 
For better comparison with Ref.~\cite{vonManteuffel:2015gxa}, we 
set the scale to $-1$ and the prefactor to $(\Gamma(d/2-1))^4$. 
Note that a factor $(\text{i}\pi^{d/2})^{-L}$, where $L$ is the number
of loops, 
is part of the integral measure used in \pysecdec{}, such that the
prefactor corresponds to Eq.~(2.4) of Ref.~\cite{vonManteuffel:2015gxa}.

The result using the \qmc{} integrator with {\tt minn=35*10**5}, {\tt minm=64}, {\tt maxeval=1},
{\tt cudablocks=128}, {\tt cudathreadsperblock=64}, {\tt maxnperpackage=8}, \\ {\tt maxmperpackage=8},
{\tt verbosity=3}, {\tt transform=`baker'}, \\ {\tt fitfunction=`polysingular'}
\begin{align}
\begin{split}
F^{\text{num.}}= &+ (3.1807379885 \pm 9.19\cdot10^{-8}) \\
                 &+ (46.10430477 \pm 1.34\cdot 10^{-6}) \cdot \epsilon \\
                 &+ \mathcal{O}(\eps^2)\;,
\end{split}
\end{align}
which can be compared to the analytical result of Ref.~\cite{vonManteuffel:2015gxa}
\begin{align}
F^{\text{analyt.}}=3.1807380843134699650 + 46.104303262308462367 \eps + \mathcal{O}(\eps^2)\,.
\end{align}
To achieve an approximate $1/n$ scaling behaviour, the Baker 
transform had to be applied to the integrand. For this 11-dimensional parameter integral, the Baker transform is superior
to the Korobov transform as it does not increase the variance of the
integrand. For details we refer to Ref.~\cite {Jahn:2018gnp}.

\subsubsection{2-loop Nbox}

\begin{figure}[h]
  \centering
\begin{subfigure}[b]{0.49\textwidth}
    \label{subfig:hfkka}\includegraphics[width=\textwidth]{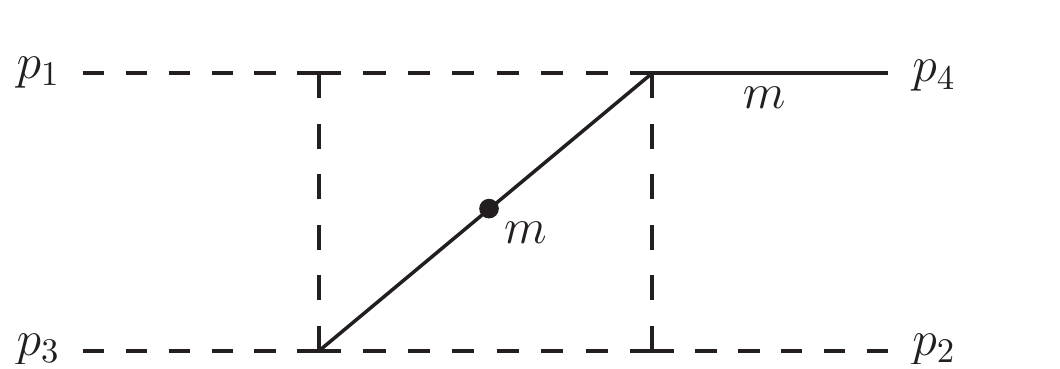}
\caption{}
\end{subfigure}
\begin{subfigure}[b]{0.49\textwidth}
    \label{subfig:hfkkbc}\includegraphics[width=\textwidth]{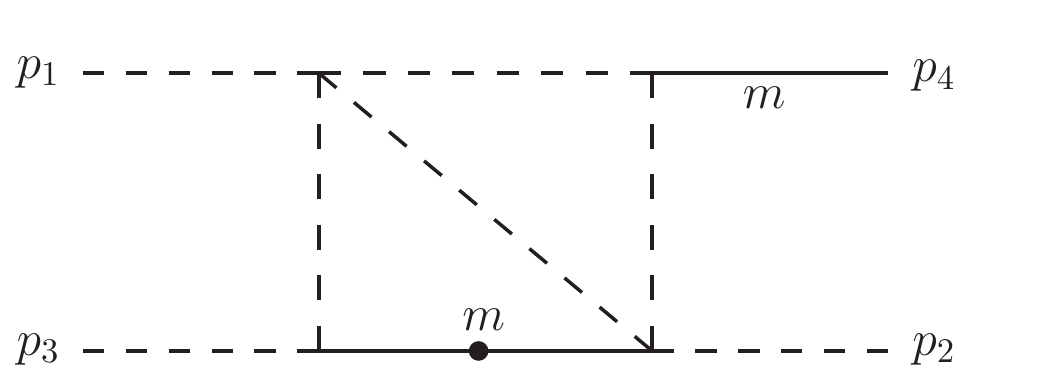}
\caption{}
\end{subfigure}
  \caption{2-loop four-point integrals with one massive propagator and one massive leg.}
  \label{fig:hfkk}
\end{figure} 
The Nbox example\footnote{Inspired by a private communication with H.~Frellesvig and K.~Kudashkin.} 
consists of three integrals, \texttt{Nbox2L\_split\_a}, \texttt{Nbox2L\_split\_b}, and \texttt{Nbox2L\_split\_c},
all of which have one massive internal line that matches the mass of one external 
leg. The integral \texttt{Nbox2L\_split\_a}  is shown in Fig.~\ref{fig:hfkk}(a), \texttt{Nbox2L\_split\_b} is 
represented in Fig.~\ref{fig:hfkk}(b) and \texttt{Nbox2L\_split\_c} by Fig.~\ref{fig:hfkk}(b) with the dot removed. 
These integrals are of interest since they have no Euclidean region and thus the sector decomposition algorithms implemented in \pysecdec{} are not guaranteed to succeed. In practice, the integral \texttt{Nbox2L\_split\_a} and \texttt{Nbox2L\_split\_b} can be computed using the {\tt split=True} option
of \pysecdec{}. The integral \texttt{Nbox2L\_split\_c} is quasi-finite and therefore does not need {\tt split=True}.

The result for the point $s=(p_1 + p_2)^2=-1$, $t=(p_1+p_3)^2=-0.8$ and $m^2=0.1$  is
\begin{align}
\begin{split}
H_a= &+ (1.54320987654321673\cdot10^{-1} \pm7.04\cdot10^{-16} \\
          &+ \text{i}\,(3.95\cdot10^{-18}\pm1.41\cdot10^{-16})) \cdot \eps^{-3} \\
     &+ (-2.6079701365346328\pm1.18\cdot10^{-14} \\
     & + \text{i}\,(1.93925472443813307\pm7.96\cdot10^{-15})) \cdot \eps^{-2} \\
     &+ (-3.73711324653151\pm1.82\cdot10^{-12}\\
     &- \text{i}\,(9.93265048220209\pm1.13\cdot10^{-12})) \cdot \eps^{-1} \\
     &+ 36.882907731123\pm2.42\cdot10^{-10} \\
     &- \text{i}\,(27.77041218391\pm8.59\cdot10^{-10})\\
     &+ \mathcal{O}(\eps)\;,
\end{split}
\end{align}
\begin{align}
\begin{split}
H_b= &+ (-8.1789971643514132\pm4.96\cdot10^{-14} \\
      &- \text{i}\,(1.71\cdot10^{-15}\pm2.80\cdot10^{-14})) \cdot \eps^{-2} \\
     &+ (-3.0495945501\pm7.22\cdot10^{-8}\\
     & - \text{i}\,(51.3901546473\pm6.58\cdot10^{-8})) \cdot \eps^{-1} \\
     &+ 160.02687326\pm2.83\cdot10^{-6} \\
     &- \text{i}\,(134.42897220\pm2.82\cdot10^{-6}) \\
     &+ \mathcal{O}(\eps)\;,
\end{split}
\end{align}
\begin{align}
\begin{split}
H_c= &+ 2.4083471021928\pm4.33\cdot10^{-11} \\
     &- \text{i}\,(25.8748336621213\pm4.59\cdot10^{-11}) \\
     &+ \mathcal{O}(\eps)\;.
\end{split}
\end{align}
To produce these results we have used the settings:
\begin{itemize}
\item {\tt minn=10**7, maxeval=1}, {\tt transform=`korobov4'}, {\tt fitfunction=`polysingular'} for $H_a$, 
\item {\tt minn=10**9, maxeval=1}, {\tt transform=`korobov6'}, {\tt fitfunction=`polysingular'} for $H_b$ and 
\item {\tt minn=15173222401}, {\tt maxeval=1}, {\tt transform=`korobov6'}, \\ {\tt generatingvectors=`cbcpt\_cfftw1\_6'} for $H_c$.
\end{itemize}

\subsubsection{6-loop bubble}

\begin{figure}[htb]
  \centering
  \includegraphics[width=0.3\textwidth]{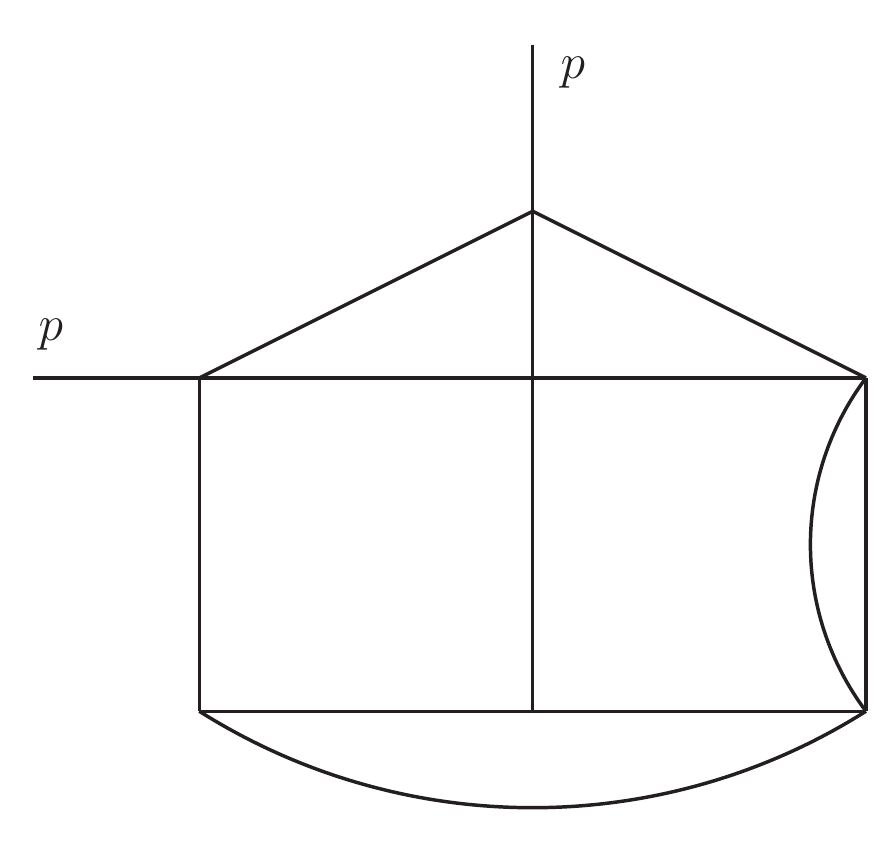}
  \caption{A 6-loop two-point integral from Ref.~\cite{Kompaniets:2017yct}.}
  \label{fig:6Lbubble}
\end{figure} 

The \texttt{bubble6L} example consists of the 6-loop 2-point integral shown in Fig.~\ref{fig:6Lbubble}. The pole coefficients are given 
analytically in Eq.~(A3) of Ref.~\cite{Kompaniets:2017yct} (at $p^2=-p_E^2=-1$, where $p_E$ is the external momentum in Euclidean space).
The \pysecdec{} symmetry finder reduces the number of sectors from more than 14000 to 8774.
We also note that the decomposition method {\tt `geometric'} needs to be used, as the method {\tt `iterative'} leads to an infinite recursion.
The analytic result is given by
\begin{align}
B_{6L}^{\text{analyt.}}&= \frac{1}{\eps^2}\,\frac{147}{16}\,\zeta_7 
- \frac{1}{\eps}\,\left(\frac{147}{16}\,\zeta_7 +\frac{27}{2}\,\zeta_3\zeta_5+\frac{27}{10}\zeta_{3,5}-\frac{2063}{504000}\,\pi^8\right) \;+\;\mathcal{O}(\epsilon^0)\nn\\
&= \frac{9.264208985946416}{\eps^2} + \frac{91.73175282208716}{\eps}  \;+\;\mathcal{O}(\epsilon^0) \;.
\end{align}

\medskip

The \pysecdec{} result at $p^2=-1$ obtained with the \qmc{} integrator using {\tt minn=10**7}, {\tt maxeval=1}, {\tt transform=`baker'},  {\tt fitfunction=`polysingular'} reads
\begin{align}
\begin{split}
B_{6L}^{\text{num.}}= 
&+(9.2642089624\pm  1.58\cdot 10^{-8})\cdot\eps^{-2} \\
& + (91.73175426 \pm 2.15\cdot 10^{-6})\cdot\eps^{-1} \\ 
&+ (1118.607204\pm  1.31\cdot 10^{-4})+\mathcal{O}(\eps)\;.
\end{split}
\end{align}

\section{Profiling}
\label{sec:profiling}
\subsection{Scaling behaviour}
\begin{figure}[htb]
  \centering
  \includegraphics[width=0.49\textwidth]{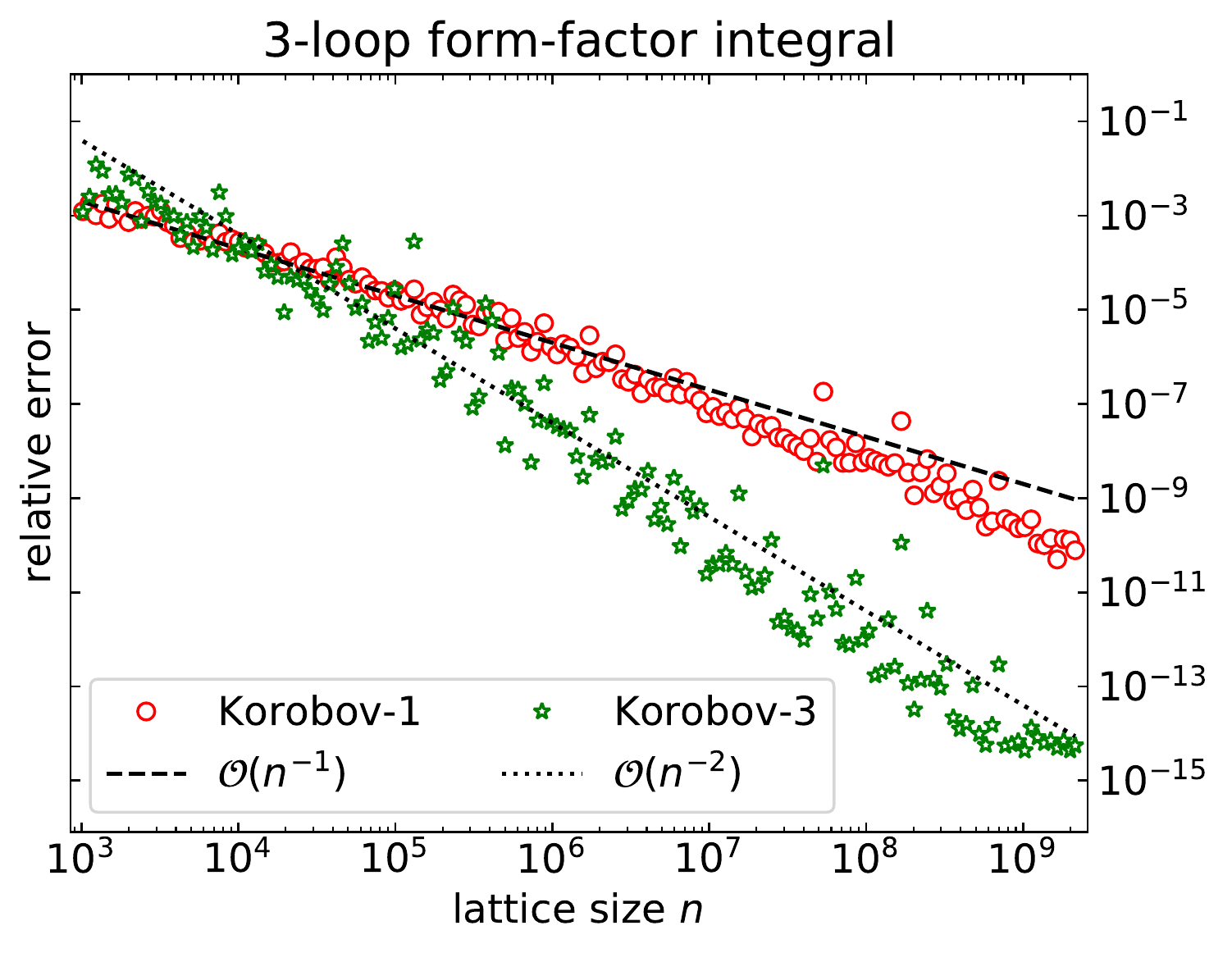}
  \includegraphics[width=0.49\textwidth]{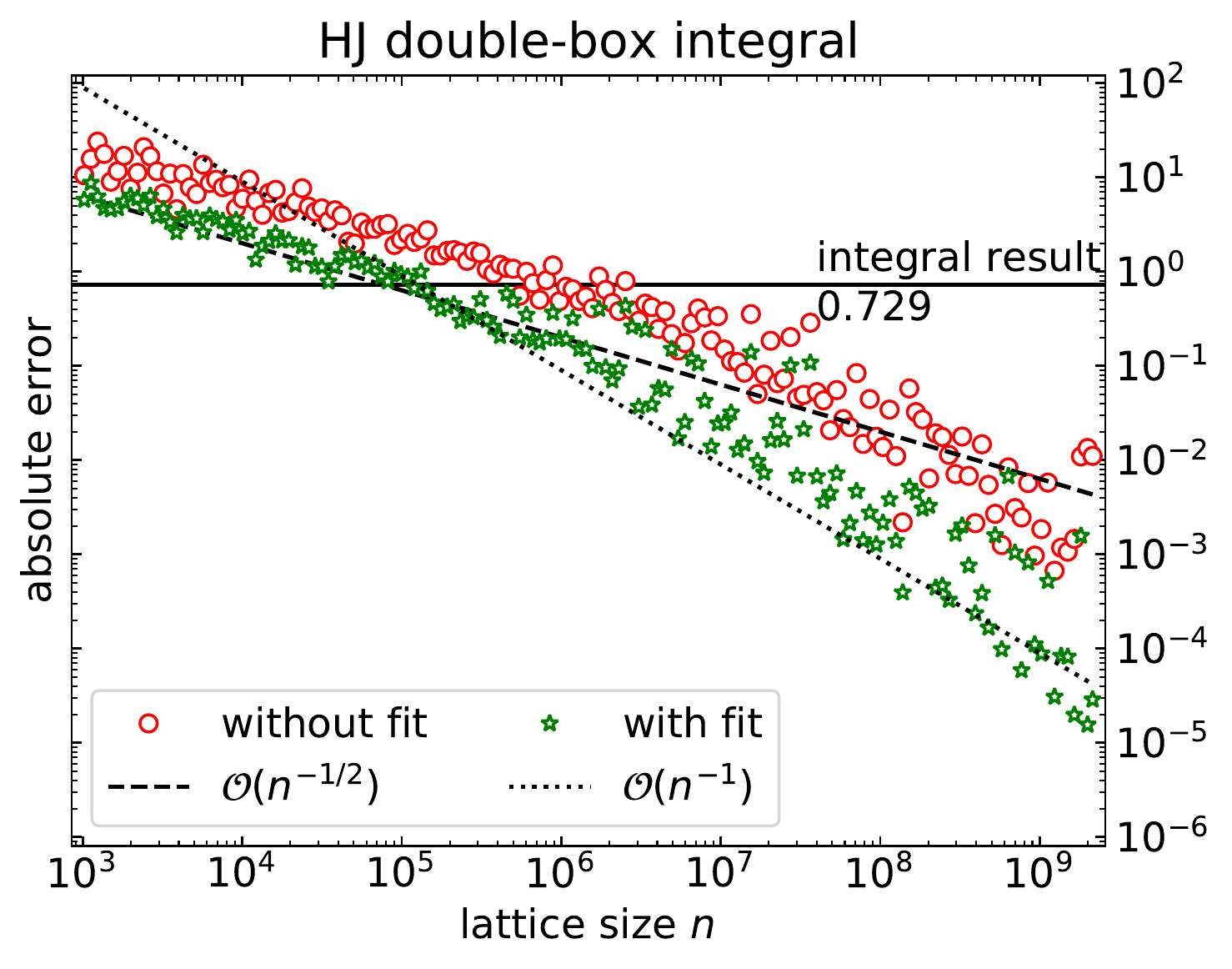}
  \caption{Scaling of the integration error with the number of lattice points $n$ for two different integrals.
  The left plot shows the relative error of the $\mathcal O(\varepsilon^4)$ contribution of a 3-loop form-factor integral using Korobov transformations of different weight.
  The plot on the right-hand side shows the scaling behaviour of a single sector of an integral, which appeared in an early stage of the calculation in Ref.~\cite{Jones:2018hbb}, also demonstrating the effect of importance sampling.}
  \label{fig:scalingplot}
\end{figure} 

Fig.~\ref{fig:scalingplot} shows how the  integration error of the \qmc{} algorithm scales with the lattice size $n$ for two different integrals.
The plot on the left-hand side shows results for the $\mathcal
O(\varepsilon^4)$ contribution of a 3-loop massless form-factor integral, which can be found in \texttt{examples/triangle3L} of the \pysecdec{} distribution. 
Using the  Korobov transformation with weight $\alpha=1$ for the periodization~(see section~\ref{sec:transformations}), we obtain per-mille-level precision for $n=1021$ and the integration error scales approximately as $\mathcal O(n^{-1})$, leading to a relative precision of $10^{-9}$ for $n\approx 10^9$. With a weight parameter $\alpha=3$, we obtain slightly larger errors for small $n$, but due to a scaling with approximately $\mathcal O(n^{-2})$, a relative precision of $10^{-14}$ can be reached with $n\approx 10^9$. We note that the expected $\mathcal O(n^{-3})$ asymptotic scaling is not observed for lattices with $n \approx 10^8$ and that, due to the use of double precision arithmetic, the integration error does not decrease when choosing even larger lattice sizes.
The plot also shows that increasing the lattice size does not always lead to a corresponding improvement of the integration error. Instead, for individual lattices, the integration error can be significantly larger than that obtained from a lattice of similar size. We observe this effect for nearly all integrals, but which lattices lead to relatively large uncertainties depends on the integrand.

The right-hand plot of Fig.~\ref{fig:scalingplot}  shows the results of an integral contributing to the NLO QCD corrections in Higgs+jet production~\cite{Jones:2018hbb} and has been selected as an example showing only slow convergence of the integration. The integrand is a single sector of a loop integral evaluated at a phase-space point with large invariant mass $m_{Hj}=8.8\,m_t$ of the Higgs-jet system, using a Korobov transform with $\alpha=1$ for the periodization. The code can be found in \texttt{examples/103\_hj\_double\_box.cpp} of the \qmc{} library. For lattice sizes $n\lesssim 10^6$, we observe that the integration error only scales with $n^{-1/2}$ and is larger than the true result of the integral. For larger lattice sizes, however, we find the expected $\mathcal{O}(n^{-1})$ scaling of the integration, allowing us to obtain the result with a precision better than $0.1\%$. 
Combining the \qmc{} with importance sampling, the integration error for small lattice sizes is reduced by about a factor of 3 and improvements by more than a factor of 10 can be seen for large  $n$.
This example shows that sampling the integrand with a lattice of sufficient size is required to obtain the desired scaling of the \qmc{} integration.  
We want to point out that for loop integrals it is possible to change the basis of required integrals using integration-by-part identities~\cite{Tkachov:1981wb,Chetyrkin:1981qh}. In many cases, this allows one to find a basis of integrals with  an improved convergence of the numerical integration. For the results presented in Ref.~\cite{Jones:2018hbb}, the integral discussed above was not used and, instead, it was possible to find  an integral basis, where evaluating the corresponding integrals with $n\approx 10^6$ was sufficient to obtain accurate results.

\subsection{Timings for test functions}
For comparison of the performance of numerical integrators, 
Genz \cite{Genz1987} has introduced a test suite, consisting of 
six integrand functions with different features: \\

\begin{tabular}{ll}
1. Oscillatory 	& $f_1(\mathbf{x}) = \text{cos}(\sum_{i=1}^n c_ix_i + 2\pi w_1)$ \\
2. Product peak	& $f_2(\mathbf{x}) = \prod_{i=1}^{n} (c_i^{-2} + (x_i-w_i)^2)^{-1}$ \\
3. Corner peak 	& $f_3(\mathbf{x}) = (1+\sum_{i=1}^n c_ix_i)^{-(n+1)}$ \\
4. Gaussian     & $f_4(\mathbf{x}) = \text{exp}(-\sum_{i=1}^n c_i^2\,(x_i - w_i)^2)$ \\
5. $C^0$ Function     & $f_5(\mathbf{x}) = \text{exp}(-\sum_{i=1}^n c_i\,|x_i - w_i|)$ \\
6. Discontinuous     & $f_6(\mathbf{x}) = \biggl\{ \begin{array}{lll} 0 & \quad & \text{if } x_1>w_1 
\text{ or } x_2>w_2\text{, } \\ \text{exp}(\sum_{i=1}^n c_ix_i) & \quad &\text{otherwise .}\end{array}$ \\
\end{tabular}\\

The test functions help to identify how well an integrator handles 
oscillatory functions, multiple periodic peaks, one peak 
anywhere in the integration region, one peak at the end of 
the integration region, $C^\infty$ functions, a continuous 
function whose derivatives are not continuous and finally a 
discontinuous function.  

The integration region for all test integrands is the unit 
hypercube. The parameters $w_i$ can be chosen randomly and should 
not affect the rate of convergence as long as $0 \leq w_i \leq 1$.
On the contrary, the positive parameters $c_i>0$ should affect 
the convergence behaviour, raising the complexity of the integral 
when $||\mathbf{c}||_1$ is increased. 

We integrate each of the above functions with the parameters:
\\
\begin{tabular}{ll}
Number of dimensions: & $ s = 5,8,10$ \\
Requested relative accuracy: & $ \epsilon_\mathrm{rel} = 10^{-8} $\\
Maximum number of samples: & $ n_\mathrm{max} = 7 \times 10^8$ \\
Time limit: & $\lesssim 360$ seconds
\end{tabular}

All tests are performed
on a machine with 2 x Intel Xeon Gold 6140 CPU @ 2.30GHz CPUs 
(36 cores, 72 threads) and 4 x Nvidia Tesla V100 GPUs.
In practice, the time limit only restricts the maximum number of samples
used by the \suave{} integrator of \cuba{} to $\lesssim 8 \times 10^7$ samples. 
Without the time limit the \suave{} integrator would take up to $3000$ seconds for some examples.

The integrand difficulties are set in accordance with Ref.~\cite{Hahn:2004fe} to:
\\
\begin{tabular}{l|l|l|l|l|l|l}
Integrand family $j$      & 1   & 2    & 3   & 4    & 5    & 6    \\
\hline
$||\mathbf{c}||_1$                & 6.0 & 18.0 & 2.2 & 15.2 & 16.1 & 16.4
\end{tabular}

For profiling we integrate each function $10$ times in each dimension, setting the values of $w_i$ and $c_i$ randomly for each integration, and average the number of correct digits obtained and time taken.
For all examples the \qmc{} is instantiated with a weight $3$ Korobov transform and the \cuba{} settings are set according to the test suite demo distributed with the latest version of \cuba{} (as of writing \cuba{} \texttt{4.2}).

\begin{table}[]
\begin{center}
\begin{tabular}{l|llllll}
 Family        & \qmc{} (CPU)       & \qmc{}  & \vegas{}     & \suave{}          & \cuhre{}    \\ \hline
 1 $(d=5)$	& 9  (0.74) & 9  (0.71) & 6	(290) & 5	(260) & 9	(2.1) \\ 
 1 $(d=8)$ 	& 8  (29) & 8  (1) & 6	(300) & 4	(270) & 9	(2.1) \\ 
 1 $(d=10)$	& 6  (68) & 6  (0.81) & 6	(340) & 4	(290) & 9	(15) \\ 
 2 $(d=5)$ 	& 10 (3.6) & 10 (0.6) & 8	(320) & 5	(280) & 9	(26) \\ 
 2 $(d=8)$ 	& 6  (63) & 6  (0.79) & 7	(330) & 5	(290) & 8	(350) \\ 
 2 $(d=10)$	& 5  (53) & 5  (0.84) & 7	(340) & 5	(290) & 8	(180) \\ 
 3 $(d=5)$ 	& 11 (3.4) & 11 (0.49) & 7	(330) & 5	(280) & 8	(15) \\ 
 3 $(d=8)$ 	& 7  (56) & 7  (0.75) & 6	(340) & 5	(290) & 9	(21) \\ 
 3 $(d=10)$	& 6  (75) & 6  (0.78) & 6	(350) & 4	(300) & 8	(92) \\ 
 4 $(d=5)$ 	& 10 (3.8) & 10 (0.63) & 6	(330) & 6	(280) & 9	(110) \\  
 4 $(d=8)$ 	& 6  (53) & 6  (0.75) & 6	(330) & 5	(290) & 9	(82) \\ 
 4 $(d=10)$	& 5  (68) & 5  (0.78) & 6	(340) & 5	(290) & 9	(140) \\  
 5 $(d=5)$ 	& 8  (35) & 8  (0.77) & 8	(320) & 5	(280) & 6	(3000) \\  
 5 $(d=8)$ 	& 5  (57) & 5  (0.74) & 7	(340) & 5	(290) & 4	(530) \\  
 5 $(d=10)$	& 4  (71) & 4  (0.79) & 7	(340) & 5	(290) & 3	(190) \\  
 6 $(d=5)$ 	& 5  (35) & 5  (0.78) & 4	(320) & 3	(280) & 4	(16) \\  
 6 $(d=8)$ 	& 4  (63) & 4  (0.72) & 4	(340) & 2	(290) & 5	(73) \\  
 6 $(d=10)$	& 3  (66) & 3  (0.76) & 5	(340) & 2	(290) & 6	(60) \\   \hline
\end{tabular}
\end{center}
\caption{Number of correct digits computed (time in seconds) for the evaluation of the test integrands 
using the \qmc{} and the integrators implemented in \cuba{} (\vegas{}, \suave{} and \cuhre{}).}
\label{tab:genztimings}
\end{table}

In Tab.~\ref{tab:genztimings} we show for each integrator the average number of digits obtained (calculated by
comparing to the analytic result) and the time taken. Note that the 
\cuba{} integrators and the \qmc{} (CPU) instance do not
make use of the GPUs, whilst the integrator denoted \qmc{} makes use of
all CPUs and GPUs. Some of \cuba{}'s algorithms sample
the integrand serially for at least part of the numerical integration,
which can greatly increase the time required to reach $n_\mathrm{max}$
evaluations. On the contrary, the \qmc{} always samples in parallel and
can usually make good use of all cores and devices. Furthermore, with the settings
suggested by the test suite demo (distributed with \cuba{}), we find that our test bed
machine is not well loaded by the \cuba{} integrators.
The load produced by the \vegas{} algorithm of \cuba{} can be
increased by up to a factor of $10$ by increasing the \texttt{nstart} and/or \texttt{nincrease} 
settings and altering the \texttt{nbatch} setting. 

The timings for the \divonne{} integrator of
\cuba{} are omitted from Tab.~\ref{tab:genztimings}. The \divonne{} algorithm, as implemented in \cuba{}, consists of 3 phases:
1) partitioning of the integration region, 2) sampling of the subregions and 3) refinement and resampling of the subregions.
With the setting $ \epsilon_\mathrm{rel} = 10^{-8} $
and a time limit of $\lesssim 360$ seconds the \divonne{} integrator usually does not complete the first phase
and so does not enter the second phase.
Without the second phase the \divonne{} integrator often underestimates the integration error and for the tests described 
in this section it typically returns results with only 2 correct digits.
The \divonne{} integrator performs much more reliably with the setting $\epsilon_\mathrm{rel} = 10^{-5}$,
returning results accurate to $4-5$ digits in around $20$ seconds in all cases.

We reiterate the warning given in Ref.~\cite{Hahn:2004fe} that the 
comparison chart should be interpreted with care. In particular, we 
emphasise that the test integrands appearing in the test suite, by virtue of 
their simplicity, bear few similarities to integrands for which 
numerical integration is typically applied. For this suite of functions the \cuhre{} routine as implemented in
\cuba{} performs best, this is due to the particular functions chosen
for the test suite and is usually not the case when applying the
integrators to sector decomposed functions. The \qmc{} performs
reasonably on the example functions, often beating the number of
digits obtained by any of the other \cuba{} integrators and taking
less time. The \qmc{} performs worse, as expected, when integrating
functions which are not smooth, in particular, the $C^0$ and
discontinuous functions. When utilising GPUs the \qmc{} typically takes
around 1 second to compute the samples (compared to $30-80$ seconds
for the CPU) regardless of the actual number of function evaluations. 
This indicates that the number of samples to be computed in these examples is too small to fully saturate the GPUs.

\begin{table}[]
\begin{center}
\begin{tabular}{l|llllll}
Family $(d=10)$ & 1 & 2 & 3 & 4 & 5 & 6 \\ \hline
Digits & 8 & 8 & 6 & 8 & 7 & 5 \\ \hline
\end{tabular}
\end{center}
\caption{Number of correct digits computed for the evaluation of the test integrands in $d=10$ 
using the \qmc{} with the Baker transform.}
\label{tab:genzbaker}
\end{table}

Relatively few correct digits are obtained by the \qmc{} when integrating the examples with $d=10$.
One reason for this is the use of the weight $3$ Korobov transform, which increases the variance of the integrand
as described at the end of Section~\ref{sec:transformations}. In Tab.~\ref{tab:genzbaker} we show 
the number of correct digits obtained using the \qmc{} with the Baker transform (rather than the Korobov transform) 
and leaving all other settings unaltered.
We observe that the number of digits obtained with the Baker transform in $d=10$ can exceed even the number of 
digits obtained in $d=8$ with the weight $3$ Korobov transform.

The source code of the program \texttt{1000\_genz\_demo}, used to perform the profiling presented in this section, is included in the {\tt examples} folder of the stand-alone \qmc{} distribution.

\subsection{Timings for loop integrals}
\begin{table}[htb]
\begin{center} 
\begin{tabular}{|l|c|c|c|c|c|c|}
\hline
& \multicolumn{2}{|c|}{\qmc{} on GPUs} & \multicolumn{2}{|c|}{\qmc{} on CPUs} & \multicolumn{2}{|c|}{\vegas{}} \\
& rel. acc. & time\,(s) & rel. acc. & time\,(s) & rel. acc. & time\,(s) \\
\hline
\scriptsize \texttt{banana 3mass 3L}      & $3.8\cdot10^{-11}$ & $15$ & $3.8\cdot10^{-11}$ & $23$ & $1.5\cdot10^{-3}$ & $39$ \\
\scriptsize \texttt{HZ nonplanar 2L}      & $1.3\cdot10^{-3}$ & $24$ & $2.1\cdot10^{-3}$ & $28$ & $5.2\cdot10^{-3}$ & $27$ \\
\scriptsize \texttt{pentabox fin 2L}      & $1.9\cdot10^{-4}$ & $42$ & $1.1\cdot10^{-3}$ & $133$ & $2.6\cdot10^{-3}$ & $139$ \\
\scriptsize \texttt{elliptic 2L}          & $2.0\cdot10^{-6}$ & $9$ & $1.6\cdot10^{-6}$ & $33$ & $3.6\cdot10^{-4}$ & $104$ \\
\scriptsize \texttt{formfactor 4L}        & $4.2\cdot10^{-7}$ & $258$ & $1.2\cdot10^{-5}$ & $235$ & $2.7\cdot10^{-4}$ & $986$ \\
\scriptsize \texttt{Nbox split b 2L}      & $2.5\cdot10^{-3}$ & $60$ & $3.5\cdot10^{-2}$ & $77$ & $1.6\cdot10^{-1}$ & $177$ \\
\scriptsize \texttt{bubble 6L}            & $8.5\cdot10^{-7}$ & $279$ & $1.1\cdot10^{-5}$ & $200$ & $5.7\cdot10^{-4}$ & $199$ \\
\hline
\end{tabular}
\end{center}
\caption{Comparison of timings using the \qmc{} on 
CPUs \& GPUs, the \qmc{} CPUs only and \vegas{} as 
implemented in the \textsc{Cuba} library.
The obtained relative accuracy refers to the finite real part
of the integral including all prefactors mentioned in 
Section~\ref{sec:examplespysd}.}
\label{tab:timings_pysd}
\end{table}

In Tab.~\ref{tab:timings_pysd}, the timings for several of the 
examples described in Section~\ref{sec:pysecdec} are compared using the \qmc{} on 
CPUs \& GPUs, the \qmc{} on CPUs and \vegas{} as 
implemented in the \cuba{} library. The timings are performed
on a machine with 2 x Intel Xeon Gold 6140 CPU @ 2.30GHz CPUs 
(36 cores, 72 threads) and 4 x Nvidia Tesla V100 GPUs.
The times reported in Tab.~\ref{tab:timings_pysd} correspond to the
wall clock times for running the integration via the python interface
of \pysecdec{}. In particular, the numerical integration of \emph{all} orders
reported for the examples given in Section~\ref{sec:examplespysd} is included in the timings.
The integrands are summed before integration (\texttt{together=True}).

The timings given in Tab.~\ref{tab:timings_pysd}
are obtained with the same parameters of the \qmc{} as stated in Section~\ref{sec:examplespysd} except for the number of samples (\texttt{minn}). The \texttt{maxeval} parameter
is set to $1$ such that the \qmc{} does not iterate. \vegas{}
is also run with a fixed number of function evaluations (\texttt{maxeval}) while the error
goals \texttt{epsrel} and \texttt{epsabs} are set to $10^{-100}$ such that they do not trigger.
The real and the imaginary part are integrated separately with \vegas{}
(\texttt{real\_complex\_together=False}).

A special situation is encountered when integrating the 4-loop form factor with
\vegas{}. The first output in verbose mode (\texttt{flags=2}) is printed to the
screen only after about fifteen minutes. We suspect this long startup time is due to the rather
large (879MB) size of the dynamic pylink library in combination with the parallelisation
using \texttt{fork} as implemented in the \cuba{} integrators library.

It is generally faster to obtain many significant digits with the \qmc{} integrator than
with the \vegas{} integrator, especially when GPUs are available. For low-precision results
however, \vegas{} can sometimes be faster.

\vspace*{3mm}

\section{Conclusions}
\label{sec:conclusion}
We have presented a quasi-Monte Carlo integrator (\qmc{}) which 
can be used both with GPUs and CPUs as a stand-alone library or within the \pysecdec{} program. 
We have described the implementation of the \qmc{}, based on a rank-1 shifted lattice rule, and given various examples of its usage.
The examples of the use of the \qmc{} within \pysecdec{} comprise a 2-loop pentagon integral, integrals which are known to contain elliptic or hyperelliptic functions, a 4-loop form factor integral and a 6-loop 2-point function.
The new version of \pysecdec{} also contains other new features, for example an improved algorithm to detect sector symmetries.

We have presented a novel approach to combine the \qmc{} integration with importance sampling.
We have investigated how the ${\cal O}(1/n)$ scaling of the error estimate depends on the dimension and form of the integrand, in particular on the transformation used to achieve a periodic integrand.
In agreement with Refs.~\cite{Li:2015foa,deDoncker:2018nqe}, we have demonstrated that rank-1 shifted lattice rules can considerably outperform integrators based on the Monte Carlo method. We also confirm that, in many cases, the use of GPUs (rather than CPUs) can lead to a speed-up of an order of magnitude or more.
This implies that the number of accurate digits which can be computed in a reasonable amount of time using our implementation is often beyond that which can be reached using \vegas{}-like Monte Carlo integration.
It should be noted, however, that the functions produced by sector decomposition are typically continuous and smooth enough to achieve ${\cal O}(1/n)$ scaling, while this is not necessarily the case for other integrands, as they occur for example in NNLO phase space integrals based on analytic subtraction of doubly unresolved real radiation. 

We believe that the method presented here, along with the easy-to-use, publicly available implementation, can boost the numerical evaluation of multi-loop amplitudes with several mass scales to an unprecedented level of automation, speed and accuracy.

The stand-alone version of the \qmc{} integrator is publicly available at\\ \url{https://github.com/mppmu/qmc}.
The new version of \pysecdec{} is available at \url{https://github.com/mppmu/secdec/releases} and 
the online documentation can be found at \url{https://secdec.readthedocs.io}.

\section*{Acknowledgements}
We would like to thank Tom Zirke for collaboration on previous
versions of the code and Oliver Schulz for providing us access and support concerning the GPU usage.
This research was supported in part by the COST Action CA16201 (`Particleface') of the European Union,
and by the Swiss National Science Foundation (SNF) under grant number 200020-175595.
SB gratefully acknowledges financial support by the ERC Starting Grant ``MathAm" (39568).
The research of JS was supported by the European Union through the ERC Advanced Grant MC@NNLO (340983).

\renewcommand \thesection{\Alph{section}}
\appendix
\setcounter{section}{0}
\setcounter{equation}{0}

\section{API documentation}
\label{sec:appendix:api}

The \qmc{} class has 7 template parameters:
\begin{itemize}
\item \texttt{T} the return type of the  function to be integrated (assumed to be a real or complex floating point type)
\item \texttt{D} the argument type of the function to be integrated (assumed to be a floating point type)
\item \texttt{M} the maximum number of integration variables of any integrand that will be passed to the integrator
\item \texttt{P} an integral transform to be applied to the integrand before integration
\item \texttt{F} a function to be fitted to the inverse cumulative distribution function of the integrand in each dimension, used to reduce the variance of the integrand (default: \texttt{fitfunctions::None::template type})
\item \texttt{G} a \cppeleven{}  style pseudo-random number engine (default: \texttt{std::mt19937\_64})
\item \texttt{H} a \cppeleven{} style uniform real distribution \\
(default: \texttt{std::uniform\_real\_distribution<D>})
\end{itemize}
Internally, unsigned integers are assumed to be of type \texttt{U = unsigned long long int}.

Typically the return type \texttt{T} and argument type \texttt{D}  are set to type double (for real numbers), \texttt{std::complex<double>} (for complex numbers on the CPU only) or \texttt{thrust::complex<double>} (for complex numbers on the GPU and CPU). In principle, the \qmc{} library supports integrating other floating point types (e.g. quadruple precision, arbitrary precision, etc), though they must be compatible with the relevant STL library functions or provide compatible overloads.

To integrate alternative floating point types, first include the header(s) defining the new type into your project and set the template arguments of the \text{} class \texttt{T} and \texttt{D} to your type. The following standard library functions must be compatible with your type or a compatible overload must be provided:
\begin{itemize}
\item {\tt sqrt}, {\tt abs}, {\tt modf}, {\tt pow}
\item {\tt std::max}, {\tt std::min}
\end{itemize}

If your type is not intended to represent a real or complex type number then you may also need to overload functions required for calculating the error resulting from the numerical integration, see the files {\tt src/overloads/real.hpp} and {\tt src/overloads/complex.hpp}. 

Example {\tt 9\_boost\_minimal\_demo} demonstrates how to instantiate the \qmc{} with a non-standard type\\ 
({\tt boost::multiprecision::cpp\_bin\_float\_quad}). To compile this example you will need the {\tt boost} library available on your system.

\subsection{Public fields}

\begin{description}

\item[Logger logger]

A wrapped \texttt{std::ostream} object to which log output from the library is written.

To write the text output of the library to a particular file, first \texttt{\#include <fstream>}, create a \texttt{std::ofstream} instance pointing to your file then set the logger of the integrator to the \texttt{std::ofstream}. For example to output very detailed output to the file \texttt{myoutput.log}:

\begin{lstlisting}
std::ofstream out_file("myoutput.log");
integrators::Qmc<double,double,MAXVAR,integrators::transforms::Korobov<3>::type> integrator;
integrator.verbosity=3;
integrator.logger = out_file;
\end{lstlisting}

\noindent Default: \texttt{std::cout}.
\\

\item[G randomgenerator]

A \cppeleven{} style pseudo-random number engine.

The seed of the pseudo-random number engine can be changed via the seed member function of the pseudo-random number engine. For total reproducibility you may also want to set \texttt{cputhreads = 1} and \texttt{devices = \{-1\}} which disables multi-threading, this helps to ensure that the floating point operations are done in the same order each time the code is run. For example:

\begin{lstlisting}
integrators::Qmc<double,double,MAXVAR,integrators::transforms::Korobov<3>::type> integrator;
integrator.randomgenerator.seed(1) // seed = 1
integrator.cputhreads = 1; // no multi-threading
integrator.devices = {-1}; // cpu only
\end{lstlisting}

\noindent Default: \texttt{std::mt19937\_64} seeded with a call to \texttt{std::random\_device}.
\\

\item[U minn]

The minimum lattice size that should be used for integration. If a lattice of the requested size is not available then {\tt n} will be the size of the next available lattice with at least {\tt minn} points. 

\noindent Default: {\tt 8191}.
\\

\item[U minm]

The minimum number of random shifts of the lattice {\tt m} that should be used to estimate the error of the result. Typically 10 to 50. 

\noindent Default: {\tt 32}.
\\

\item[D epsrel]

The relative error that the \qmc{} should attempt to achieve. 

\noindent Default: {\tt 0.01}.
\\

\item[D epsabs]

The absolute error that the \qmc{} should attempt to achieve. For real types the integrator tries to find an estimate {\tt E} for the integral {\tt I} 
which fulfills  {\tt |E-I| <= max(epsabs, epsrel*I)}. For complex types the goal is controlled by the {\tt errormode} setting.

\noindent Default: {\tt 1e-7}.
\\

\item[U maxeval]

The (approximate) maximum number of function evaluations that should be performed while integrating. The actual number of function evaluations can be slightly larger if there is not a suitably sized lattice available. 

\noindent Default: {\tt 1000000}.
\\

\item[U maxnperpackage]

Maximum number of points to compute per thread per work package. 

\noindent Default: {\tt 1}.
\\

\item[U maxmperpackage]

Maximum number of shifts to compute per thread per work package. 

\noindent Default: {\tt 1024}.
\\

\item[ErrorMode errormode]

Controls the error goal that the library attempts to achieve when the integrand return type is a complex type. For real types the {\tt errormode} setting is ignored.
Possible values:
\begin{itemize}
\item  {\tt all} - try to find an estimate {\tt E} for the integral {\tt I} which fulfills\\
  {\tt |E-I| <= max(epsabs, epsrel*I)} for each component (real and imaginary) separately,
\item   {\tt largest} - try to find an estimate {\tt E} for the integral {\tt I} such that\\ 
$ \mathrm{max}( |Re[E]-Re[I]|, |Im[E]-Im[I]| ) \leq \mathrm{max}( \eps_{\rm{abs}}, \eps_{\rm{rel}}\cdot \mathrm{max}( |Re[I]|,|Im[I]| ) )$, 
i.e. to achieve either the {\tt epsabs} error goal or that the largest error is smaller than {\tt epsrel} times the value of the largest component (either real or imaginary).
\end{itemize}
\noindent Default: {\tt all}.
\\

\item[U cputhreads]

The number of CPU threads that should be used to evaluate the integrand function. If GPUs are used 1 additional CPU thread per device will be launched for communicating with the device. 

\noindent Default: {\tt std::thread::hardware\_concurrency()}.
\\

\item[U cudablocks]

The number of blocks to be launched on each \cuda{} device. 

\noindent Default: (determined at run time).
\\

\item[U cudathreadsperblock]

The number of threads per block to be launched on each \cuda{} device. \cuda{} kernels launched by the \qmc{} library have the execution configuration {\tt <<< cudablocks, cudathreadsperblock >>>}. For more information on how to optimally configure these parameters for your hardware and/or integral refer to the Nvidia guidelines. 

\noindent Default: (determined at run time).
\\

\item[{\tt std::set<int> devices}]

A set of devices on which the integrand function should be evaluated. The device id {\tt -1} represents all CPUs present on the system, the field {\tt cputhreads} can be used to control the number of CPU threads spawned. The indices {\tt 0,1,...} are device ids of \cuda{} devices present on the system. 

\noindent Default: {\tt {-1,0,1,...,nd}} where {\tt nd} is the number of \cuda{} devices detected on the system.
\\

\item[{\tt std::map<U,std::vector<U>> generatingvectors}]

A map of available generating vectors which can be used to generate a lattice. The implemented \qmc{} algorithm requires that the generating vectors be generated with a prime lattice size. By default the library uses generating vectors with 100 components, thus it supports integration of functions with up to 100 dimensions.
The default generating vectors have been generated with lattice size chosen as the next prime number above $(110/100)^i\cdot 1020$ for {\tt i} between {\tt 0} and {\tt 152}, additionally the lattice $2^{31}-1$ ({\tt INT\_MAX} for {\tt int32}) is included. 

\noindent Default: {\tt cbcpt\_dn1\_100()}.
\\
\item[U evaluateminn]

The minimum lattice size that should be used by the \texttt{evaluate} function to evaluate the integrand, if variance reduction is enabled these points are used for fitting the inverse cumulative distribution function. If a lattice of the requested size is not available then n will be the size of the next available lattice with at least \texttt{evaluateminn} points.

\noindent Default: \texttt{100000}.
\\

\item[U verbosity]

Possible values: {\tt 0,1,2,3}. Controls the verbosity of the output to {\tt logger} of the \qmc{} library.
\begin{itemize}
\item[0] - no output,
\item[1] - key status updates and statistics,
\item[2] - detailed output, useful for debugging,
\item[3] - very detailed output, useful for debugging.
\end{itemize}
\noindent Default: {\tt 0}.
\\

\item[size\_t fitstepsize]

Controls the number of points included in the fit used for variance reduction. A step size of \texttt{x} includes (after sorting by value) every \texttt{x}th point in the fit.

\noindent Default: \texttt{10}.
\\

\item[size\_t fitmaxiter]

See \texttt{maxiter} in the non-linear least-squares fitting \texttt{GSL} documentation.

\noindent Default: \texttt{40}.
\\

\item[double fitxtol]

See \texttt{xtol} in the non-linear least-squares fitting \texttt{GSL} documentation.

\noindent Default: \texttt{3e-3}.
\\

\item[double fitgtol]

See \texttt{gtol} in the non-linear least-squares fitting \texttt{GSL} documentation.

\noindent Default: \texttt{1e-8}.
\\

\item[double fitftol]

See \texttt{ftol} in the non-linear least-squares fitting \texttt{GSL} documentation.

\noindent Default: \texttt{1e-8}.
\\

\item[gsl\_multifit\_nlinear\_parameters fitparametersgsl]\hspace*{2cm}

\noindent See \texttt{gsl\_multifit\_nlinear\_parameters} in the non-linear least-squares fitting \texttt{GSL} documentation.

\noindent Default: \texttt{\{\}}.

\end{description}

\subsection{Public Member Functions}

\begin{description}

\item[U get\_next\_n(U preferred\_n)]

Returns the lattice size $n$ of the lattice in {\tt generatingvectors} that is greater than or equal to {\tt preferred\_n}. This represents the size of the lattice that would be used for integration if {\tt minn} was set to {\tt preferred\_n}.
\\

\item[\texttt{template <typename I> result<T,U> integrate(I\& func)}]

Integrates the function {\tt func} in $d$ dimensions using the integral transform {\tt transform}. The result is returned in a {\tt result} struct with the following members:
\begin{itemize}
\item \texttt{integral} - the result of the integral
\item \texttt{error} - the estimated absolute error of the result
\item \texttt{n} - the size of the largest lattice used during integration
\item \texttt{m} - the number of shifts of the largest lattice used during integration.
\item \texttt{U iterations} - the number of iterations used during integration
\item \texttt{U evaluations} - the total number of function evaluations during integration
\end{itemize}

The functor \texttt{func} must define its dimension as a public member variable \texttt{number\_of\_integration\_variables}.

\noindent Calls: \texttt{get\_next\_n}.
\\

\item[\texttt{template <typename I> samples<T,D> evaluate(I\& func)}]

Evaluates the functor \texttt{func} on a lattice of size greater than or equal to \texttt{evaluateminn}. The samples are returned in a \texttt{samples} struct with the following members:
\begin{itemize}
\item \texttt{std::vector<U> z} - the generating vector of the lattice used to produce the samples
\item \texttt{std::vector<D> d} - the random shift vector used to produce the samples
\item \texttt{std::vector<T> r} - the values of the integrand at each randomly shifted lattice point
\item \texttt{U n} - the size of the lattice used to produce the samples
\item \texttt{D get\_x(const U sample\_index, const U integration\_variable\_index)} - a function which returns the argument (specified by \texttt{integration\_variable\_index}) used to evaluate the integrand for a specific sample (specified by \texttt{sample\_index}).
\end{itemize}

The functor \texttt{func} must define its dimension as a public member variable \texttt{number\_of\_integration\_variables}.

\noindent Calls: \texttt{get\_next\_n}.
\\

\item[\texttt{template <typename I> typename F<I,D,M>::transform\_t fit(I\& func)}]

Fits a function (specified by the type \texttt{F} of the integrator) to the inverse cumulative distribution function of the integrand dimension-by-dimension and returns a functor representing the new integrand after this variance reduction procedure.

The functor \texttt{func} must define its dimension as a public member variable \texttt{number\_of\_integration\_variables}.

\noindent Calls: \texttt{get\_next\_n}, \texttt{evaluate}.

\end{description}

\subsection{Generating vectors}
\label{sec:appendix:generatingVectors}

We offer generating vectors for different lattice sizes $n$, 
and also for different maximal dimensions $s$: $s\leq 6$ or $s\leq 100$.
The generating vectors which are distributed with the version described in this paper are summarised in Table~\ref{tab:genvecs}.
We used the so-called Component-By-Component (CBC) construction~\cite{nuyens2006fast}, 
computed using partly D.~Nuyens'  {\tt fastrank1pt.m} tool~\cite{nuyensCBC} and, for very large lattice sizes, our own CBC tool based on 
the FFTW algorithm~\cite{FFTW}.

\begin{table}[htb]
\begin{tabular}{|l|c|l|l|}
\hline
 Name & Max.  & Computed via & Lattice Sizes \\
&Dimension&&\\
 \hline
 {\tt cbcpt\_dn1\_100} & 100 &  {\it fastrank1pt.m}  tool~\cite{nuyensCBC} & 1021 - 2147483647 \\
 {\tt cbcpt\_dn2\_6}     & 6     & {\it fastrank1pt.m} tool~\cite{nuyensCBC} & 65521 - 2499623531 \\
 {\tt cbcpt\_cfftw1\_6} & 6     & CBC tool based on \cite{FFTW} & 2500000001 - 15173222401 \\
 \hline
\end{tabular}
\caption{Types of generating vectors distributed with the program.\label{tab:genvecs}}
\end{table}

The generating vectors distributed with the code are produced for Korobov spaces with smoothness $\alpha=2$, in the notation of Ref~\cite{Nuyens:2016proceedings} we use:
\begin{itemize}
\item Kernel $\omega(x)=2 \pi^2 (x^2 - x + 1/6)$,
\item Weights $\gamma_i = 1/s$ for $i = 1, \ldots, s$,
\item Parameters $\beta_i = 1$ for $i = 1, \ldots, s$.
\end{itemize}

The generating vectors used by the \qmc{} can be selected by setting the integrator's {\tt generatingvectors} member variable. 
Example (assuming an integrator instance named {\tt integrator}):
\begin{lstlisting}[numbers=none]
integrator.generatingvectors = integrators::generatingvectors::cbcpt_dn2_6();
\end{lstlisting}

If you prefer to use custom generating vectors and/or 100 dimensions and/or 15173222401 lattice points is not enough, you can supply your own generating vectors. Compute your generating vectors using another tool then put them into a map and set \texttt{generatingvectors}. For example, to instruct the \qmc{} to use only two generating vectors ($\mathbf{z} = (1,3)$ for $n=7$ and $\mathbf{z} = (1,7)$  for $n=11$) the \texttt{generatingvectors} map would be set as follows:
\begin{lstlisting}
std::map<unsigned long long int,std::vector<unsigned long long int>> my_generating_vectors = { {7, {1,3}}, {11, {1,7}} };
integrators::Qmc<double,double,10> integrator;
integrator.generatingvectors = my_generating_vectors;
\end{lstlisting}

\subsection{Integral Transforms}
\label{sec:appendix:integraltransforms}

\begin{table}[htb]
\begin{tabular}{|l|l|}
\hline
 Name & Description \\
 \hline
 {\tt Korobov<r\_0,r\_1>} & A polynomial integral transform with weight $\propto x^{r_0}  (1-x)^{r_1}$ \\
 {\tt Korobov<r>} & A polynomial integral transform with weight $\propto x^r  (1-x)^r$ \\
 {\tt Sidi<r>} & A trigonometric integral transform with weight $\propto \sin^r(\pi x)$ \\
 {\tt Baker} & The baker's transformation, $\phi(x) = 1 - |2x-1|$ \\
 {\tt None} & The trivial transform, $\phi(x) = x$ \\
 \hline
\end{tabular}
\caption{Types of periodizing transformations distributed with the program.\label{tab:integraltransforms}}
\end{table}

The integral transforms distributed with the \qmc{} are listed in Table~\ref{tab:integraltransforms}.
The integral transform used by the \qmc{} can be selected when constructing the \qmc{}. 
Example (assuming a real type integrator instance named integrator):\\
\begin{lstlisting}[numbers=none]
integrators::Qmc<double,double,10,integrators::transforms::Korobov<5,3>::type> integrator;
\end{lstlisting}
instantiates an integrator which applies a weight $(r_0=5,r_1=3)$ Korobov transform to the integrand before integration.

\subsection{Fit Functions}
\label{sec:appendix:fitfunctions}

\begin{table}[htb]
\begin{tabular}{|l|l|}
\hline
 Name & Description \\
 \hline
 {\tt PolySingular} & A \nth{3} order polynomial with two additional $1/(p-x)$ terms, \\
 & \vspace{-2em}\parbox{0em}{\begin{align*}
 f(x) = &  \frac{|p_2|(x (p_0-1))}{(p_0-x)} + \frac{|p_3| (x (p_1-1))}{(p_1-x)}  \\
 & + x (p_4+x (p_5+x (1-|p_2|-|p_3|-p_4-p_5))) \\
 \end{align*}} \\
 {\tt None} & The trivial transform, $f(x) = x$ \\
 \hline
\end{tabular}
\caption{Types of fit functions distributed with the program.\label{tab:fitfunctions}}
\end{table}

The fit function used by the \qmc{} can be selected when constructing the \qmc{}. These functions are used to approximate the inverse cumulative distribution function of the integrand dimension-by-dimension. Example (assuming a real type integrator instance named \texttt{integrator}):
\begin{lstlisting}[numbers=none]
integrators::Qmc<double,double,10,integrators::transforms::Korobov<3>::type,integrators::fitfunctions::PolySingular::type> integrator;
\end{lstlisting}
instantiates an integrator which reduces the variance of the integrand by fitting a \texttt{PolySingular} type function before integration.
Possible fit functions are given in Table~\ref{tab:fitfunctions}.

\bibliographystyle{JHEP}


\begin{thebibliography}{10}

\bibitem{Laporta:2004rb}
S.~Laporta and E.~Remiddi, \emph{{Analytic treatment of the two loop equal mass
  sunrise graph}},
  \href{http://dx.doi.org/10.1016/j.nuclphysb.2004.10.044}{\emph{Nucl. Phys.}
  {\bfseries B704} (2005) 349--386},
  [\href{https://arxiv.org/abs/hep-ph/0406160}{{\ttfamily hep-ph/0406160}}].

\bibitem{Adams:2017tga}
L.~Adams, E.~Chaubey and S.~Weinzierl, \emph{{Simplifying Differential
  Equations for Multiscale Feynman Integrals beyond Multiple Polylogarithms}},
  \href{http://dx.doi.org/10.1103/PhysRevLett.118.141602}{\emph{Phys. Rev.
  Lett.} {\bfseries 118} (2017) 141602},
  [\href{https://arxiv.org/abs/1702.04279}{{\ttfamily 1702.04279}}].

\bibitem{Abreu:2017enx}
S.~Abreu, R.~Britto, C.~Duhr and E.~Gardi, \emph{{Algebraic Structure of Cut
  Feynman Integrals and the Diagrammatic Coaction}},
  \href{http://dx.doi.org/10.1103/PhysRevLett.119.051601}{\emph{Phys. Rev.
  Lett.} {\bfseries 119} (2017) 051601},
  [\href{https://arxiv.org/abs/1703.05064}{{\ttfamily 1703.05064}}].

\bibitem{Primo:2017ipr}
A.~Primo and L.~Tancredi, \emph{{Maximal cuts and differential equations for
  Feynman integrals. An application to the three-loop massive banana graph}},
  \href{http://dx.doi.org/10.1016/j.nuclphysb.2017.05.018}{\emph{Nucl. Phys.}
  {\bfseries B921} (2017) 316--356},
  [\href{https://arxiv.org/abs/1704.05465}{{\ttfamily 1704.05465}}].

\bibitem{Bourjaily:2017bsb}
J.~L. Bourjaily, A.~J. McLeod, M.~Spradlin, M.~von Hippel and M.~Wilhelm,
  \emph{{Elliptic Double-Box Integrals: Massless Scattering Amplitudes beyond
  Polylogarithms}},
  \href{http://dx.doi.org/10.1103/PhysRevLett.120.121603}{\emph{Phys. Rev.
  Lett.} {\bfseries 120} (2018) 121603},
  [\href{https://arxiv.org/abs/1712.02785}{{\ttfamily 1712.02785}}].

\bibitem{Broedel:2017kkb}
J.~Broedel, C.~Duhr, F.~Dulat and L.~Tancredi, \emph{{Elliptic polylogarithms
  and iterated integrals on elliptic curves I: general formalism}},
  \href{http://dx.doi.org/10.1007/JHEP05(2018)093}{\emph{JHEP} {\bfseries 05}
  (2018) 093}, [\href{https://arxiv.org/abs/1712.07089}{{\ttfamily
  1712.07089}}].

\bibitem{Adams:2018yfj}
L.~Adams and S.~Weinzierl, \emph{{The $\varepsilon$-form of the differential
  equations for Feynman integrals in the elliptic case}},
  \href{http://dx.doi.org/10.1016/j.physletb.2018.04.002}{\emph{Phys. Lett.}
  {\bfseries B781} (2018) 270--278},
  [\href{https://arxiv.org/abs/1802.05020}{{\ttfamily 1802.05020}}].

\bibitem{Broedel:2018iwv}
J.~Broedel, C.~Duhr, F.~Dulat, B.~Penante and L.~Tancredi, \emph{{Elliptic
  symbol calculus: from elliptic polylogarithms to iterated integrals of
  Eisenstein series}},
  \href{http://dx.doi.org/10.1007/JHEP08(2018)014}{\emph{JHEP} {\bfseries 08}
  (2018) 014}, [\href{https://arxiv.org/abs/1803.10256}{{\ttfamily
  1803.10256}}].

\bibitem{Lee:2018ojn}
R.~N. Lee, A.~V. Smirnov and V.~A. Smirnov, \emph{{Evaluating ‘elliptic’
  master integrals at special kinematic values: using differential equations
  and their solutions via expansions near singular points}},
  \href{http://dx.doi.org/10.1007/JHEP07(2018)102}{\emph{JHEP} {\bfseries 07}
  (2018) 102}, [\href{https://arxiv.org/abs/1805.00227}{{\ttfamily
  1805.00227}}].

\bibitem{Bourjaily:2018ycu}
J.~L. Bourjaily, Y.-H. He, A.~J. Mcleod, M.~Von~Hippel and M.~Wilhelm,
  \emph{{Traintracks through Calabi-Yau Manifolds: Scattering Amplitudes beyond
  Elliptic Polylogarithms}},
  \href{http://dx.doi.org/10.1103/PhysRevLett.121.071603}{\emph{Phys. Rev.
  Lett.} {\bfseries 121} (2018) 071603},
  [\href{https://arxiv.org/abs/1805.09326}{{\ttfamily 1805.09326}}].

\bibitem{Blumlein:2018cms}
J.~Bl{\"u}mlein and C.~Schneider, \emph{{Analytic computing methods for
  precision calculations in quantum field theory}},
  \href{http://dx.doi.org/10.1142/S0217751X18300156}{\emph{Int. J. Mod. Phys.}
  {\bfseries A33} (2018) 1830015},
  [\href{https://arxiv.org/abs/1809.02889}{{\ttfamily 1809.02889}}].

\bibitem{Broedel:2018qkq}
J.~Broedel, C.~Duhr, F.~Dulat, B.~Penante and L.~Tancredi, \emph{{Elliptic
  Feynman integrals and pure functions}},
  \href{http://dx.doi.org/10.1007/JHEP01(2019)023}{\emph{JHEP} {\bfseries 01}
  (2019) 023}, [\href{https://arxiv.org/abs/1809.10698}{{\ttfamily
  1809.10698}}].

\bibitem{Bourjaily:2018yfy}
J.~L. Bourjaily, A.~J. McLeod, M.~von Hippel and M.~Wilhelm, \emph{{Bounded
  Collection of Feynman Integral Calabi-Yau Geometries}},
  \href{http://dx.doi.org/10.1103/PhysRevLett.122.031601}{\emph{Phys. Rev.
  Lett.} {\bfseries 122} (2019) 031601},
  [\href{https://arxiv.org/abs/1810.07689}{{\ttfamily 1810.07689}}].

\bibitem{Hepp:1966eg}
K.~Hepp, \emph{Proof of the {B}ogolyubov-{P}arasiuk theorem on
  renormalization}, {\emph{Commun. Math. Phys.} {\bfseries 2} (1966) 301--326}.

\bibitem{Roth:1996pd}
M.~Roth and A.~Denner, \emph{High-energy approximation of one-loop {F}eynman
  integrals}, {\emph{Nucl. Phys.} {\bfseries B479} (1996) 495--514},
  [\href{https://arxiv.org/abs/hep-ph/9605420}{{\ttfamily hep-ph/9605420}}].

\bibitem{Binoth:2000ps}
T.~Binoth and G.~Heinrich, \emph{An automatized algorithm to compute infrared
  divergent multi-loop integrals}, {\emph{Nucl. Phys.} {\bfseries B585} (2000)
  741--759}, [\href{https://arxiv.org/abs/hep-ph/0004013}{{\ttfamily
  hep-ph/0004013}}].

\bibitem{Heinrich:2008si}
G.~Heinrich, \emph{{Sector Decomposition}},
  \href{http://dx.doi.org/10.1142/S0217751X08040263}{\emph{Int. J. Mod. Phys.}
  {\bfseries A23} (2008) 1457--1486},
  [\href{https://arxiv.org/abs/0803.4177}{{\ttfamily 0803.4177}}].

\bibitem{Becker:2012bi}
S.~Becker and S.~Weinzierl, \emph{{Direct numerical integration for multi-loop
  integrals}},
  \href{http://dx.doi.org/10.1140/epjc/s10052-013-2321-1}{\emph{Eur. Phys. J.}
  {\bfseries C73} (2013) 2321},
  [\href{https://arxiv.org/abs/1211.0509}{{\ttfamily 1211.0509}}].

\bibitem{Sborlini:2016hat}
G.~F.~R. Sborlini, F.~Driencourt-Mangin and G.~Rodrigo, \emph{{Four-dimensional
  unsubtraction with massive particles}},
  \href{http://dx.doi.org/10.1007/JHEP10(2016)162}{\emph{JHEP} {\bfseries 10}
  (2016) 162}, [\href{https://arxiv.org/abs/1608.01584}{{\ttfamily
  1608.01584}}].

\bibitem{Freitas:2016sty}
A.~Freitas, \emph{{Numerical multi-loop integrals and applications}},
  \href{http://dx.doi.org/10.1016/j.ppnp.2016.06.004}{\emph{Prog. Part. Nucl.
  Phys.} {\bfseries 90} (2016) 201--240},
  [\href{https://arxiv.org/abs/1604.00406}{{\ttfamily 1604.00406}}].

\bibitem{deDoncker:2017gnb}
E.~de~Doncker, F.~Yuasa, K.~Kato, T.~Ishikawa, J.~Kapenga and O.~Olagbemi,
  \emph{{Regularization with Numerical Extrapolation for Finite and
  UV-Divergent Multi-loop Integrals}},
  \href{http://dx.doi.org/10.1016/j.cpc.2017.11.001}{\emph{Comput. Phys.
  Commun.} {\bfseries 224} (2018) 164--185},
  [\href{https://arxiv.org/abs/1702.04904}{{\ttfamily 1702.04904}}].

\bibitem{Gluza:2016fwh}
J.~Gluza, T.~Jelinski and D.~A. Kosower, \emph{{Efficient Evaluation of Massive
  Mellin-Barnes Integrals}},
  \href{http://dx.doi.org/10.1103/PhysRevD.95.076016}{\emph{Phys. Rev.}
  {\bfseries D95} (2017) 076016},
  [\href{https://arxiv.org/abs/1609.09111}{{\ttfamily 1609.09111}}].

\bibitem{Usovitsch:2018shx}
J.~Usovitsch, I.~Dubovyk and T.~Riemann, \emph{{MBnumerics: Numerical
  integration of Mellin-Barnes integrals in physical regions}}, {\emph{PoS}
  {\bfseries LL2018} (2018) 046},
  [\href{https://arxiv.org/abs/1810.04580}{{\ttfamily 1810.04580}}].

\bibitem{Baglio:2018lrj}
J.~Baglio, F.~Campanario, S.~Glaus, M.~M{\"u}hlleitner, M.~Spira and
  J.~Streicher, \emph{{Gluon fusion into Higgs pairs at NLO QCD and the top
  mass scheme}},
  \href{http://dx.doi.org/10.1140/epjc/s10052-019-6973-3}{\emph{Eur. Phys. J.}
  {\bfseries C79} (2019) 459},
  [\href{https://arxiv.org/abs/1811.05692}{{\ttfamily 1811.05692}}].

\bibitem{Bendavid:2018nar}
\emph{{Les Houches 2017: Physics at TeV Colliders Standard Model Working Group
  Report}}, 2018.

\bibitem{Blondel:2018mad}
A.~Blondel et~al., \emph{{Standard Model Theory for the FCC-ee: The Tera-Z}},
  in \emph{{Mini Workshop on Precision EW and QCD Calculations for the FCC
  Studies: Methods and Techniques; CERN, Geneva, Switzerland, January 12-13,
  2018}}, 2018.
\newblock \href{https://arxiv.org/abs/1809.01830}{{\ttfamily 1809.01830}}.

\bibitem{Bogner:2007cr}
C.~Bogner and S.~Weinzierl, \emph{{Resolution of singularities for multi-loop
  integrals}},
  \href{http://dx.doi.org/10.1016/j.cpc.2007.11.012}{\emph{Comput.Phys.Commun.}
  {\bfseries 178} (2008) 596--610},
  [\href{https://arxiv.org/abs/0709.4092}{{\ttfamily 0709.4092}}].

\bibitem{Gluza:2010rn}
J.~Gluza, K.~Kajda, T.~Riemann and V.~Yundin, \emph{{Numerical Evaluation of
  Tensor Feynman Integrals in Euclidean Kinematics}},
  \href{http://dx.doi.org/10.1140/epjc/s10052-010-1516-y}{\emph{Eur.Phys.J.}
  {\bfseries C71} (2011) 1516},
  [\href{https://arxiv.org/abs/1010.1667}{{\ttfamily 1010.1667}}].

\bibitem{Ueda:2009xx}
T.~Ueda and J.~Fujimoto, \emph{{New implementation of the sector decomposition
  on FORM}}, {\emph{PoS} {\bfseries ACAT08} (2008) 120},
  [\href{https://arxiv.org/abs/0902.2656}{{\ttfamily 0902.2656}}].

\bibitem{Smirnov:2008py}
A.~Smirnov and M.~Tentyukov, \emph{{Feynman Integral Evaluation by a Sector
  decomposiTion Approach (FIESTA)}},
  \href{http://dx.doi.org/10.1016/j.cpc.2008.11.006}{\emph{Comput.Phys.Commun.}
  {\bfseries 180} (2009) 735--746},
  [\href{https://arxiv.org/abs/0807.4129}{{\ttfamily 0807.4129}}].

\bibitem{Smirnov:2009pb}
A.~Smirnov, V.~Smirnov and M.~Tentyukov, \emph{{FIESTA 2: Parallelizeable
  multiloop numerical calculations}},
  \href{http://dx.doi.org/10.1016/j.cpc.2010.11.025}{\emph{Comput.Phys.Commun.}
  {\bfseries 182} (2011) 790--803},
  [\href{https://arxiv.org/abs/0912.0158}{{\ttfamily 0912.0158}}].

\bibitem{Smirnov:2013eza}
A.~V. Smirnov, \emph{{FIESTA 3: cluster-parallelizable multiloop numerical
  calculations in physical regions}},
  \href{http://dx.doi.org/10.1016/j.cpc.2014.03.015}{\emph{Comput.Phys.Commun.}
  {\bfseries 185} (2014) 2090--2100},
  [\href{https://arxiv.org/abs/1312.3186}{{\ttfamily 1312.3186}}].

\bibitem{Smirnov:2015mct}
A.~V. Smirnov, \emph{{FIESTA4: Optimized Feynman integral calculations with GPU
  support}}, \href{http://dx.doi.org/10.1016/j.cpc.2016.03.013}{\emph{Comput.
  Phys. Commun.} {\bfseries 204} (2016) 189--199},
  [\href{https://arxiv.org/abs/1511.03614}{{\ttfamily 1511.03614}}].

\bibitem{Borowka:2015mxa}
S.~Borowka, G.~Heinrich, S.~P. Jones, M.~Kerner, J.~Schlenk and T.~Zirke,
  \emph{{SecDec-3.0: numerical evaluation of multi-scale integrals beyond one
  loop}}, \href{http://dx.doi.org/10.1016/j.cpc.2015.05.022}{\emph{Comput.
  Phys. Commun.} {\bfseries 196} (2015) 470--491},
  [\href{https://arxiv.org/abs/1502.06595}{{\ttfamily 1502.06595}}].

\bibitem{Borowka:2017idc}
S.~Borowka, G.~Heinrich, S.~Jahn, S.~P. Jones, M.~Kerner, J.~Schlenk et~al.,
  \emph{{pySecDec: a toolbox for the numerical evaluation of multi-scale
  integrals}}, \href{http://dx.doi.org/10.1016/j.cpc.2017.09.015}{\emph{Comput.
  Phys. Commun.} {\bfseries 222} (2018) 313--326},
  [\href{https://arxiv.org/abs/1703.09692}{{\ttfamily 1703.09692}}].

\bibitem{Borowka:2018dsa}
S.~Borowka, T.~Gehrmann and D.~Hulme, \emph{{Systematic approximation of
  multi-scale Feynman integrals}},
  \href{http://dx.doi.org/10.1007/JHEP08(2018)111}{\emph{JHEP} {\bfseries 08}
  (2018) 111}, [\href{https://arxiv.org/abs/1804.06824}{{\ttfamily
  1804.06824}}].

\bibitem{Hahn:2004fe}
T.~Hahn, \emph{{CUBA: A library for multidimensional numerical integration}},
  \href{http://dx.doi.org/10.1016/j.cpc.2005.01.010}{\emph{Comput. Phys.
  Commun.} {\bfseries 168} (2005) 78--95},
  [\href{https://arxiv.org/abs/hep-ph/0404043}{{\ttfamily hep-ph/0404043}}].

\bibitem{Hahn:2014fua}
T.~Hahn, \emph{{Concurrent Cuba}},
  \href{https://arxiv.org/abs/1408.6373}{{\ttfamily 1408.6373}}.

\bibitem{QMCActaNumerica}
J.~Dick, F.~Y. Kuo and I.~H. Sloan, \emph{High-dimensional integration: The
  quasi-monte carlo way}, {\emph{Acta Numerica} {\bfseries 22} (2013)
  133--288}.

\bibitem{Li:2015foa}
Z.~Li, J.~Wang, Q.-S. Yan and X.~Zhao, \emph{{Efficient Numerical Evaluation of
  Feynman Integral}},
  \href{http://dx.doi.org/10.1088/1674-1137/40/3/033103}{\emph{{Chinese Physics
  C}} {\bfseries 40, No. 3} (2016) 033103},
  [\href{https://arxiv.org/abs/1508.02512}{{\ttfamily 1508.02512}}].

\bibitem{deDoncker:2018nqe}
E.~de~Doncker, A.~Almulihi and F.~Yuasa, \emph{{High-speed evaluation of loop
  integrals using lattice rules}},
  \href{http://dx.doi.org/10.1088/1742-6596/1085/5/052005}{\emph{J. Phys. Conf.
  Ser.} {\bfseries 1085} (2018) 052005}.

\bibitem{Borowka:2016ehy}
S.~Borowka, N.~Greiner, G.~Heinrich, S.~Jones, M.~Kerner, J.~Schlenk et~al.,
  \emph{{Higgs Boson Pair Production in Gluon Fusion at Next-to-Leading Order
  with Full Top-Quark Mass Dependence}},
  \href{http://dx.doi.org/10.1103/PhysRevLett.117.079901,
  10.1103/PhysRevLett.117.012001}{\emph{Phys. Rev. Lett.} {\bfseries 117}
  (2016) 012001}, [\href{https://arxiv.org/abs/1604.06447}{{\ttfamily
  1604.06447}}].

\bibitem{Borowka:2016ypz}
S.~Borowka, N.~Greiner, G.~Heinrich, S.~P. Jones, M.~Kerner, J.~Schlenk et~al.,
  \emph{{Full top quark mass dependence in Higgs boson pair production at
  NLO}}, \href{http://dx.doi.org/10.1007/JHEP10(2016)107}{\emph{JHEP}
  {\bfseries 10} (2016) 107},
  [\href{https://arxiv.org/abs/1608.04798}{{\ttfamily 1608.04798}}].

\bibitem{Jones:2018hbb}
S.~P. Jones, M.~Kerner and G.~Luisoni, \emph{{Next-to-Leading-Order QCD
  Corrections to Higgs Boson Plus Jet Production with Full Top-Quark Mass
  Dependence}},
  \href{http://dx.doi.org/10.1103/PhysRevLett.120.162001}{\emph{Phys. Rev.
  Lett.} {\bfseries 120} (2018) 162001},
  [\href{https://arxiv.org/abs/1802.00349}{{\ttfamily 1802.00349}}].

\bibitem{KuoNuyensPractical}
F.~Y. Kuo and D.~Nuyens, ``Lecture notes: A practical guide to quasi-monte
  carlo methods.'' National Chiao Tung University \& National Taiwan
  University, November, 2016.

\bibitem{SLOAN19981}
I.~H. Sloan and H.~Woźniakowski, \emph{When are quasi-monte carlo algorithms
  efficient for high dimensional integrals?},
  \href{http://dx.doi.org/https://doi.org/10.1006/jcom.1997.0463}{\emph{Journal
  of Complexity} {\bfseries 14} (1998) 1 -- 33}.

\bibitem{nuyens2006fast}
D.~Nuyens and R.~Cools, \emph{Fast algorithms for component-by-component
  construction of rank-1 lattice rules in shift-invariant reproducing kernel
  hilbert spaces}, {\emph{Mathematics of Computation} {\bfseries 75} (2006)
  903--920}.

\bibitem{Korobov1963}
N.~M. Korobov, \emph{Number-theoretic methods in approximate analysis},
  {\emph{Fizmatgiz} {\bfseries Moscow} (1963) }.

\bibitem{LAURIE1996337}
D.~P. Laurie, \emph{Periodizing transformations for numerical integration},
  \href{http://dx.doi.org/https://doi.org/10.1016/0377-0427(95)00196-4}{\emph{Journal
  of Computational and Applied Mathematics} {\bfseries 66} (1996) 337 -- 344}.

\bibitem{Kuo2007}
F.~Y. Kuo, I.~H. Sloan and H.~Wo{\'{z}}niakowski, \emph{Periodization strategy
  may fail in high dimensions},
  \href{http://dx.doi.org/10.1007/s11075-007-9145-8}{\emph{Numerical
  Algorithms} {\bfseries 46} (Dec, 2007) 369--391}.

\bibitem{Sidi1993}
A.~Sidi, \emph{A New Variable Transformation for Numerical Integration}.
\newblock Birkh{\"a}user Basel, Basel, 1993.

\bibitem{baker_trafo}
F.~J. Hickernell, \emph{{Obtaining $O(N^{-2+\epsilon})$ convergence for lattice
  quadrature rules, In: K.-T. Fang, F. J. Hickernell, H. Niederreiter (eds.)
  Monte Carlo and Quasi-Monte Carlo Methods 2000}}.
\newblock Springer, Berlin, 2002.

\bibitem{Lepage:1977sw}
G.~P. Lepage, \emph{{A New Algorithm for Adaptive Multidimensional
  Integration}}, \href{http://dx.doi.org/10.1016/0021-9991(78)90004-9}{\emph{J.
  Comput. Phys.} {\bfseries 27} (1978) 192}.

\bibitem{Gough:2009:GSL:1538674}
M.~Galassi and al., \emph{GNU Scientific Library Reference Manual - Third
  Edition}.
\newblock Network Theory Ltd., 3rd~ed., 2009.

\bibitem{Bonciani:2016qxi}
R.~Bonciani, V.~Del~Duca, H.~Frellesvig, J.~M. Henn, F.~Moriello and V.~A.
  Smirnov, \emph{{Two-loop planar master integrals for Higgs$\to 3$ partons
  with full heavy-quark mass dependence}},
  \href{http://dx.doi.org/10.1007/JHEP12(2016)096}{\emph{JHEP} {\bfseries 12}
  (2016) 096}, [\href{https://arxiv.org/abs/1609.06685}{{\ttfamily
  1609.06685}}].

\bibitem{Georgoudis:2015hca}
A.~Georgoudis and Y.~Zhang, \emph{{Two-loop Integral Reduction from Elliptic
  and Hyperelliptic Curves}},
  \href{http://dx.doi.org/10.1007/JHEP12(2015)086}{\emph{JHEP} {\bfseries 12}
  (2015) 086}, [\href{https://arxiv.org/abs/1507.06310}{{\ttfamily
  1507.06310}}].

\bibitem{vonManteuffel:2015gxa}
A.~von Manteuffel, E.~Panzer and R.~M. Schabinger, \emph{{On the Computation of
  Form Factors in Massless QCD with Finite Master Integrals}},
  \href{http://dx.doi.org/10.1103/PhysRevD.93.125014}{\emph{Phys. Rev.}
  {\bfseries D93} (2016) 125014},
  [\href{https://arxiv.org/abs/1510.06758}{{\ttfamily 1510.06758}}].

\bibitem{Jahn:2018gnp}
S.~Jahn, \emph{{Numerical evaluation of multi-loop integrals}},
  \href{http://dx.doi.org/10.22323/1.303.0019}{\emph{PoS} {\bfseries LL2018}
  (2018) 019}.

\bibitem{Kompaniets:2017yct}
M.~V. Kompaniets and E.~Panzer, \emph{{Minimally subtracted six loop
  renormalization of $O(n)$-symmetric $\phi^4$ theory and critical exponents}},
  \href{http://dx.doi.org/10.1103/PhysRevD.96.036016}{\emph{Phys. Rev.}
  {\bfseries D96} (2017) 036016},
  [\href{https://arxiv.org/abs/1705.06483}{{\ttfamily 1705.06483}}].

\bibitem{Tkachov:1981wb}
F.~V. Tkachov, \emph{{A Theorem on Analytical Calculability of Four Loop
  Renormalization Group Functions}},
  \href{http://dx.doi.org/10.1016/0370-2693(81)90288-4}{\emph{Phys. Lett.}
  {\bfseries 100B} (1981) 65--68}.

\bibitem{Chetyrkin:1981qh}
K.~G. Chetyrkin and F.~V. Tkachov, \emph{{Integration by Parts: The Algorithm
  to Calculate beta Functions in 4 Loops}},
  \href{http://dx.doi.org/10.1016/0550-3213(81)90199-1}{\emph{Nucl. Phys.}
  {\bfseries B192} (1981) 159--204}.

\bibitem{Genz1987}
A.~Genz, \emph{A Package for Testing Multiple Integration Subroutines}.
\newblock Springer Netherlands, Dordrecht, 1987,
  \href{http://dx.doi.org/{https://doi.org/10.1007/978-94-009-3889-2\_33}}{{https://doi.org/10.1007/978-94-009-3889-2\_33}}.

\bibitem{nuyensCBC}
D.~Nuyens. https://people.cs.kuleuven.be/~dirk.nuyens/fast-cbc/.

\bibitem{FFTW}
M.~Frigo, S.~Johnson and et. al. https://github.com/FFTW/fftw3.

\bibitem{Nuyens:2016proceedings}
D.~Nuyens and R.~Cools, \emph{Fast component-by-component construction, a
  reprise for different kernels},  in \emph{Monte Carlo and Quasi-Monte Carlo
  Methods 2004} (H.~Niederreiter and D.~Talay, eds.), (Berlin, Heidelberg),
  pp.~373--387, Springer Berlin Heidelberg, 2006.

\end{thebibliography}

\providecommand{\href}[2]{#2}\begingroup\raggedright\endgroup

\end{document}